\documentclass[12pt,a4paper,oneside]{article}

\usepackage{geometry}
\usepackage{color}
\usepackage[centertags]{amsmath}
\usepackage{stmaryrd}
\usepackage{dsfont}
\usepackage{mathrsfs}
\usepackage{url}

\usepackage{hyperref}
\usepackage[utf8]{inputenc}
\usepackage{polski}
\usepackage{amsfonts,amssymb,amsmath,amscd,slashed,amsthm,mathtools}
\usepackage{enumerate}
\usepackage{graphicx}
\usepackage{verbatim}
\usepackage{tikz}
\usepackage{psfrag}
\usepackage{epsfig}

\usepackage[polish,english]{babel}

\usepackage{hyphenat}
\newtheorem{Definition}{Definition}
\newtheorem{Conjecture}{Conjecture}
\newtheorem{Example}{Example}
\newtheorem{Theorem}{Theorem}
\newtheorem{Claim}{Claim}
\newtheorem{Lemma}{Lemma}
\newtheorem{Remark}{Remark}
\newtheorem{Corollary}{Corollary}
\newtheorem{Proposition}{Proposition}

\newenvironment{Proof}{\textbf{Proof}}{ \qed \\}
\newenvironment{Thm}[1][]{\begin{Theorem} \normalfont \textbf{#1} \itshape}{\end{Theorem}}

\newenvironment{Def}[1][]{\begin{Definition} \normalfont \textbf{#1} \itshape}{\end{Definition}}

\newenvironment{Lem}[1][]{\begin{Lemma} \normalfont \textbf{#1} \itshape}{\end{Lemma}}
\newenvironment{Rem}[1][]{\begin{Remark} \normalfont \textbf{#1}}{\end{Remark}}
\newenvironment{Cor}[1][]{\begin{Corollary} \normalfont \textbf{#1} \itshape}{\end{Corollary}}
\newenvironment{Prop}[1][]{\begin{Proposition} \normalfont \textbf{#1} \itshape}{\end{Proposition}}

\newcommand{\ev}{\textnormal{ev}}
\newcommand{\supp}{\textnormal{supp}}
\newcommand{\grad}{\textnormal{grad}}
\newcommand{\id}{\textnormal{id}}
\newcommand{\sN}{{\mathbb N}}

\newcommand{\sR}{{\mathbb R}}

\newcommand{\Bf}{\mathscr{B}}

\newcommand{\cC}{\mathscr{C}}
\newcommand{\cCi}{\mathscr{C}_{\textnormal{inext}}}

\newcommand{\I}{\mathcal{I}}

\newcommand{\K}{\mathcal{K}}

\newcommand{\M}{\mathcal{M}}

\newcommand{\Pf}{\mathscr{P}}

\newcommand{\Sw}{\mathcal{S}}
\newcommand{\T}{\mathcal{T}}
\newcommand{\X}{\mathcal{X}}
\newcommand{\Y}{\mathcal{Y}}
\newcommand{\U}{\mathcal{U}}

\newcommand{\bA}{\textbf{A}}
\newcommand{\bB}{\textbf{B}}

\newcommand{\itSigma}{\mathit{\Sigma}}

\newcommand{\itPi}{\mathit{\Pi}}

\newcommand{\bsigma}{\boldsymbol\sigma}
\newcommand{\bgamma}{\boldsymbol\gamma}

\newcommand{\bpi}{\boldsymbol\pi}

\newcommand{\eps}{\varepsilon}

\begin{document}

\title{Polish spaces of causal curves}

\author{Tomasz Miller}

\date{
{\footnotesize Faculty of Mathematics and Information Science, Warsaw University of Technology}\\
{\footnotesize ul. Koszykowa 75, 00-662 Warsaw, Poland}\\[0.3cm]
{\footnotesize Copernicus Center for Interdisciplinary Studies}\\
{\footnotesize ul. Szczepa\'nska 1/5, 31-011 Krak\'ow, Poland}\\[0.3cm]
{\footnotesize T.Miller@mini.pw.edu.pl}
}

\maketitle

\begin{abstract}
We propose and study a new approach to the topologization of spaces of (possibly not all) future-directed causal curves in a stably causal spacetime. It relies on parametrizing the curves ``in accordance'' with a chosen time function. Thus obtained topological spaces of causal curves are separable and completely metrizable, i.e. Polish. The latter property renders them particularly useful in the optimal transport theory. To illustrate this fact, we explore the notion of a causal time-evolution of measures in globally hyperbolic spacetimes and discuss its physical interpretation.
\end{abstract}

MSC classes: 53C50, 53C80, 28E99, 60B05

\section{Introduction and main results}
\label{sec::intro}

The notion of a causal curve is one of the central concepts in mathematical relativity. Causal curves not only model the worldlines of physical particles, but also determine the causal structure of a given spacetime. It is therefore not surprising that the topological properties of (the particular subsets of) the set of all causal curves can provide an insight into the structure of a given spacetime. Most notably, the historically first definition of global hyperbolicity due to Leray \cite{Leray} (later adopted by Geroch \cite{Geroch1970}) invoked the compactness of the set of all causal curves linking two distinct events. Since then, various topologizations of spaces of causal curves were considered, each of them having its advantages and disadvantages. The two most popular approaches lead through the so-called $C^0$-topology \cite{Geroch1970,HELargeScaleStructure,Penrose1972}, and through the compact-open topology on the space of causal curves parametrized by their arc-length \cite{YCB67,MinguzziCurves08}. See also \cite{Samann16,SanchezProgress} for more details.

In the present paper, we propose a new approach to the topologization of spaces of (possibly not all) future-directed causal curves in a given stably causal spacetime $\M$. Concretely (cf. Definition \ref{CITDef}), for any time function $\T: \M \rightarrow \sR$ and any interval $I$ we introduce the space $C^I_\T$ of all future-directed causal curves $\gamma: I \rightarrow \M$ such that
\begin{align*}
\exists c_\gamma > 0 \ \, \forall s,t \in I \ \quad \T(\gamma(t)) - \T(\gamma(s)) = c_\gamma (t-s),
\end{align*}
endowed with the compact-open topology.

In other words, instead of parametrizing the causal curves by their arc-length, we decide to parametrize them ``in accordance'' with a chosen time function $\T$. The main advantage of this approach lies in the exceptionally good topological properties of thus obtained spaces, summarized by the adjective: Polish (cf. Proposition \ref{PropCITPolish}).

Polish spaces are separable and completely metrizable topological spaces. Their properties render them extremely useful in probability theory \cite{ModProb}, optimal transport theory \cite{UsersGuide,ambrosio2008gradient} and mathematical logic (the so-called descriptive set theory is founded on them \cite{Kechris}). Important examples of Polish spaces include second-countable locally compact Hausdorff (LCH) spaces (hence, in particular, manifolds) as well as the spaces $C(\X,\Y)$ of continuous maps from a second-countable LCH space $\X$ into a Polish space $\Y$, endowed with the compact-open topology \cite[Example A.10.]{KerrLi}. Finally, given a Polish space $\Y$, the space $\Pf(\Y)$ of all Borel probability measures on $\Y$ (endowed with the narrow topology) is Polish itself \cite[Remark 7.1.7]{ambrosio2008gradient}.

In the context of mathematical relativity, it was already Geroch who noted that the space of causal paths (i.e. images of curves) connecting two distinct events equipped with the $C^0$-topology is separable and metrizable \cite{Geroch1970}. Geroch's approach was extended by Penrose \cite{Penrose1972}, who studied the space $\cC$ of \emph{all} compact causal paths (i.e. without fixing their endpoints) endowed with the $C^0$-topology. In any case, however, the important question of completeness has not been addressed.

The article aims to fill this gap and to further study the topological properties of the spaces of causal curves and paths. Let us briefly summarize the paper's main results.

We shall write $[\gamma]$ to denote the image of the causal curve $\gamma$. We prove that for any \emph{compact} interval $I$ and any time function $\T$ the map $[ \, . \, ]: C^I_\T \rightarrow \cC$, $\gamma \mapsto [\gamma]$ is in fact a homeomorphism (Corollary \ref{CITallpathsCor} \& Theorem \ref{CIThomeo}). This implies, in particular, that the space $\cC$ is actually Polish (at least for $\M$ stably causal), and thus the spaces $C^I_\T$ with $I$ compact provide an alternative (yet mathematically equivalent) view on the $C^0$-topology.

What about the case when $I$ is a \emph{noncompact} interval? This question turns out to be much more involved and we answer it only partially, focusing mostly on the case when $I = \sR$ and $\M$ is globally hyperbolic. In particular, we prove that for any Cauchy temporal functions $\T_1,\T_2$ the reparametrization map $C^{\sR}_{\T_1} \ni \gamma \mapsto \widetilde{\gamma} \in C^{\sR}_{\T_2}$, defined via $\widetilde{\gamma} := \gamma \circ \left(\T_2 \circ \gamma \right)^{-1} \circ \T_1 \circ \gamma$, is a \emph{homeomorphism} (Theorem \ref{RhomeoThm}). What is more, for any Cauchy temporal function $\T$ the map $C^{\sR}_\T \ni \gamma \mapsto [\gamma]$ is a well-defined surjection onto the set $\cCi$ of all \emph{inextendible} causal paths (Proposition \ref{CITallpaths4}).

For the sake of turning $[ \, . \, ]$ into a bijection, we restrict it to $\I_\T := \{ \gamma \in C^\sR_\T \ | \ \T \circ \gamma = \id_\sR \}$, which is a closed (and hence Polish) subspace of $C^\sR_\T$. This suggests a way to topologize the set $\cCi$ by transporting the topology from $\I_\T$ using the now-bijective map $[ \, . \, ]$. Of course, such a topology might a priori depend on the Cauchy temporal function $\T$, which would make it rather artificial. However, this turns out not to be the case, as the reparametrization map $\gamma \mapsto \widetilde{\gamma}$ defined above yields a homeomorphism between $\I_{\T_1}$ and $\I_{\T_2}$ (Proposition \ref{IT2}).

Both introduced classes of spaces $C^\sR_\T$ and $\I_\T$ turn out to have a natural application in the Lorentzian version of the optimal transport theory. The latter is a fast developing area of research \cite{Bertrand2013,Brenier2003,Suhr2016}, which e.g. opened up a novel approach to the early universe reconstruction problem \cite{BrenierFrisch2003,Frisch2002,Frisch2006}.

In our previous work \cite{EcksteinMiller2015} we extended the standard causal precedence relation $\preceq$ between events of a given spacetime $\M$ onto the space of measures (by which term we shall always mean \emph{Borel probability measures}) on $\M$ and studied its properties. Building upon these results, in the current paper we investigate the notion of the \emph{causal time-evolution of measures} in globally hyperbolic spacetimes. Concretely, the main result of this part of the article is the following equivalence.
\begin{Thm}
\label{main}
Let $\M$ be a globally hyperbolic spacetime and let $\T: \M \rightarrow \sR$ be a Cauchy temporal function. Let also $I \subseteq \sR$ be any interval. Consider the map $\mu: I \rightarrow \Pf(\M)$, $t \mapsto \mu_t$ such that $\supp \, \mu_t \subseteq \T^{-1}(t)$ for every $t \in I$. The following conditions are equivalent:
\begin{enumerate}[\itshape i)]
\item The map $\mu$ is \emph{causal} in the sense that
\begin{align}
\label{causal_measure_map}
\forall \, s,t \in I \quad s \leq t \ \Rightarrow \ \mu_s \preceq \mu_t.
\end{align}
\item There exists $\sigma \in \Pf(C_\T^I)$ such that $(\ev_t)_\# \sigma = \mu_t$ for every $t \in I$, where $\ev_t: C_\T^I \rightarrow \M$ denotes the evaluation map.
\end{enumerate}
\end{Thm}
This theorem might be regarded as an analogue of \cite[Theorem 2.10]{UsersGuide}, in which geodesics in a Polish geodesic space are replaced with future-directed causal curves in a globally hyperbolic spacetime, or as a (distant) cousin of the result called the ``superposition principle'' as given in \cite[Theorem 3]{Bernard12} (see also \cite[Theorem 3.2]{ambrosio2008continuity} or \cite[Theorem 6.2.2]{Crippa2007PhD} for other formulations).

Last but not least, we prove also the following variant of Theorem \ref{main} for the spaces $\I_\T$.
\begin{Thm}
\label{mainvar}
Let $\M, \T$ be as above. For any map $\mu: \sR \rightarrow \Pf(\M)$, $t \mapsto \mu_t$ the following conditions are equivalent:
\begin{enumerate}[\itshape i)]
\item The map $\mu$ satisfies $\supp \, \mu_t \subseteq \T^{-1}(t)$ for every $t \in \sR$ and is causal.
\item There exists $\upsilon \in \Pf(\I_\T)$ such that $(\ev_t|_{\I_\T})_\# \upsilon = \mu_t$ for every $t \in \sR$.
\end{enumerate}
\end{Thm}

The outline of the paper is as follows: we begin in Section~\ref{sec::discussion} by discussing the physical interpretation of Theorems \ref{main} and \ref{mainvar} in the context of mathematical relativity. More concretely, we explore the phenomenon of the causal time-evolution of physical quantities distributed in space as seen by different observers. Section~\ref{sec::PSoCC} studies in detail the Polish spaces of causal curves introduced above as well as their mutual relationships. Section~\ref{sec::proof} contains the proofs of Theorems \ref{main} and \ref{mainvar}, preceded by recalling and developing some additional tools and results on the verge of Lorentzian geometry and optimal transport theory. In order to make the paper self-contained, we finish with the \hyperref[sec::preliminaries]{Appendix} containing some basic definitions and results from causality theory needed in the article.

\section{Physical discussion}
\label{sec::discussion}

Let us first discuss the motivation behind calling a measure-valued map \emph{causal} in the sense introduced above as well as the physical content of Theorems \ref{main} and \ref{mainvar}.

Suppose physicist $\bA$ wants to describe the time-evolution of a physical quantity $Q$ distributed in space --- be it a mass or charge distribution --- whose total amount is conserved. Assume first that the background spacetime is Minkowski. In order to describe a dynamical phenomenon, the physicist more or less implicitly chooses the time parameter $t$ and thus employs a particular foliation of the Minkowski spacetime by $t$-slices --- the hypersurfaces of simultaneity. In $\bA$'s description, the distribution of $Q$ at the time instant $t$ might be modelled by a measure $\mu_t \in \Pf(\M)$ (we normalize the total amount of $Q$ to one) supported on the corresponding $t$-slice.

$\bA$ is interested in whether the time-evolution $t \mapsto \mu_t$ does not violate Einstein's causality, understood as the impossibility of superluminal propagation of any physical object or interaction. In case of $Q$, which is a quantity distributed in space, the velocity bound concerns the ``infinitesimal portions'' of $Q$ (also called \emph{parcels}). Intuitively, the distribution of $Q$ evolves causally if each of its parcels travels along a future-directed causal curve.

The above intuition is made mathematically rigorous by demanding that $s \leq t$ implies $\mu_s \preceq \mu_t$, where $\preceq$ denotes the causal precedence relation between measures on $\M$. This relation, introduced and studied in full generality in \cite{EcksteinMiller2015}, extends the standard causality relation between events through the notion of a \emph{coupling} borrowed from the optimal transport theory \cite{Villani2008}, and is defined as follows.
\begin{Def}
\label{JGPcausality_def_true}
Let $\M$ be a~spacetime. For any $\mu,\nu \in \Pf(\M)$ we say that $\mu$ \emph{causally precedes} $\nu$ (symbolically $\mu \preceq \nu$) if there exists a \emph{causal coupling} of $\mu$ and $\nu$, by which we mean $\omega \in \Pf(\M^2)$ such that
\begin{enumerate}[\itshape i)]
\item $\pi^1_\# \omega = \mu$ and $\pi^2_\# \omega = \nu$,
\item $\omega(J^+) = 1$,
\end{enumerate}
where $\pi^i: \M^2 \rightarrow \M$ denotes the projection on the $i$-th argument, $i=1,2$.
\end{Def}
Let us emphasize that the left-hand side of the latter condition is well-defined, because $J^+$ is a Borel subset of $\M^2$ \cite[Section 3]{EcksteinMiller2015}. The name ``causal coupling'' was independently coined in \cite{Suhr2016}. For more concrete physical examples of causal measure-valued maps the reader is referred to \cite{2NEW2016}.

In this setting, what Theorems \ref{main} and \ref{mainvar} say is that $\bA$ can provide an alternative description of the studied dynamical phenomenon. Concretely, instead of using a causal measure-valued map $t \mapsto \mu_t$, $\bA$ can use a single probability measure $\sigma$ living on the suitable Polish space of causal curves. Both descriptions are equivalent, i.e. they contain exactly the same information.

Let us emphasize that Theorem \ref{main} works not only in the Minkowski spacetime, but in any globally hyperbolic spacetime $\M$. In order to describe the time-evolution of $Q$, the physicist $\bA$ chooses a Cauchy temporal function $\T_1$ and uses the Geroch--Bernal--S\'{a}nchez (GBS) splitting (see \hyperref[sec::preliminaries]{Appendix}, Theorem \ref{GBS}) and thus picks a particular foliation of $\M$ by the Cauchy hypersurfaces comprising the level sets of $\T_1$. In this case, every instantaneous distribution of $Q$ is modelled by a measure $\mu_t$ supported on the corresponding level set $\T_1^{-1}(t)$.

Even in the Minkowski case, however, the following questions arise: Suppose that another physicist $\bB$ would like to describe \emph{the same} dynamical phenomenon, but choosing a \emph{different} Cauchy temporal function $\T_2$ and thus a different foliation of $\M$. Of course, $\bB$ would obtain a completely different family of measures, living on Cauchy hypersurfaces transversal to those employed by $\bA$. The questions are: What is the relationship between $\bA$'s and $\bB$'s descriptions? Are there any invariants within their descriptions? Would they always agree about the causality of their measure-valued maps? Below we address these questions in a rigorous way.

Let the map $I_1 \ni t \mapsto \mu_t$, $\supp \, \mu_t \subseteq \T^{-1}_1(t)$ be the time-evolution of a physical quantity $Q$ as described by physicist $\bA$ employing the Cauchy temporal function $\T_1$. Similarly, the map $I_2 \ni \tau \mapsto \nu_\tau$, $\supp \, \nu_\tau \subseteq \T^{-1}_2(\tau)$ will denote the time-evolution of $Q$ according to the description of physicist $\bB$ who uses the Cauchy temporal function $\T_2$. Since we assume that $Q$ is conserved in the course of evolution, we take $I_1, I_2 = \sR$. Albeit fragmentary descriptions in which the dynamical parameter $t \in I_1 \subsetneq \sR$ are practically useful (cf. \cite{2NEW2016} and references therein), they unavoidably distinguish at least one Cauchy hypersurface (e.g. $\T_1^{-1}(\inf I_1)$ if $I_1$ were bounded from below) and so they are trivially not GBS-splitting-independent.

Assume that both maps $t \mapsto \mu_t$ and $\tau \mapsto \nu_\tau$ are causal. Because $I_1,I_2 = \sR$, both $\bA$ and $\bB$ can apply Theorem \ref{mainvar} and encapsulate their descriptions within the measures $\upsilon_i \in \Pf(\I_{\T_i})$, $i=1,2$, respectively. As it was stated in Section \ref{sec::intro}, we have a homeomorphism $\widetilde{\ }: \I_{\T_1} \rightarrow \I_{\T_2}$, which is a suitable reparametrization map. Let $\widetilde{\upsilon}_1$ denote the pushforward of $\upsilon_1$ by $\widetilde{\ }$. We argue that $\upsilon_2 = \widetilde{\upsilon}_1$.

The physical reasons for this equality to be true can be explained as follows. Imagine $\bA$ and $\bB$ want to describe the motion of a pointlike particle. They are given its unparametrized worldline -- a single inextendible causal path $[\gamma] \in \cCi$. Physicist $\bA$ parametrizes $[\gamma]$ by assigning to each $p \in [\gamma]$ the number $\T_1(p)$, obtaining the curve $\gamma_1 \in \I_{\T_1}$. Similarly, physicist $\bB$ obtains the curve $\gamma_2 \in \I_{\T_2}$ assigning to each $p \in [\gamma]$ the number $\tau := \T_2(p)$. We claim that $\gamma_2 = \widetilde{\gamma}_1$. Indeed, recall from Section \ref{sec::intro} that the maps $\I_{\T_i} \ni \gamma \mapsto [\gamma]_i$, $i=1,2$ are well-defined bijections onto $\cCi$ (see Proposition \ref{IT1} for the proof). We obviously have
\begin{align*}
[\gamma_2]_2 = [\gamma] = [\gamma_1]_1 = [\widetilde{\gamma}_1]_2,
\end{align*}
where the last equality is true because $\widetilde{\gamma}_1$ is a reparametrization of $\gamma_1$. The claim follows from the injectivity of $[\, . \,]_2$.

In other words, it is precisely the homeomorphism $\widetilde{\ }: \I_{\T_1} \rightarrow \I_{\T_2}$, which allows to switch between $\bA$'s and $\bB$'s parametrizations of any given inextendible causal path. Via its pushforward map, the above argumentation extends onto \emph{measures on the spaces of causal curves} and we conclude that $\upsilon_2 = \widetilde{\upsilon}_1$.

We can further exploit the above reasoning with the aid of the Polish space structure we endowed $\cCi$ with. Namely, for $i=1,2$, let $[\upsilon_i]_i \in \Pf(\cCi)$ denote the pushforward of $\upsilon_i$ by the bijection $[\, . \,]_i$, which is a homeomorphism by the very definition of the topology on $\cCi$. Because for any $\gamma_1 \in \I_{\T_1}$ we know that $[\widetilde{\gamma}_1]_2 = [\gamma_1]_1$, therefore for measures we obtain $[\upsilon_2]_2 = [\widetilde{\upsilon}_1]_2 = [\upsilon_1]_1$.

We have thus obtained an invariant implicitly contained in both $\bA$'s and $\bB$'s descriptions. It is a measure on the space $\cCi$ of inextendible causal paths, and as such it does not pertain to any particular GBS splitting. To put it differently: Just as the path $[\gamma] \in \cCi$ is the GBS-splitting-independent spatiotemporal object modelling the motion of a pointlike particle, one can analogously say that the measure $[\upsilon] \in \Pf(\cCi)$ is the GBS-splitting-independent spatiotemporal object which models the dynamics of a physical quantity $Q$ distributed in space.

With the above in mind, we now address the question whether $\bA$ and $\bB$ would always agree on the causality of their measure-valued maps. To begin with, suppose that $\bA$'s map $t \mapsto \mu_t$ is causal. Therefore, $\bA$ can rephrase his/her description in the form of $\upsilon_1 \in \Pf(\I_{\T_1})$ and then in the form of $[\upsilon_1] =: [\upsilon] \in \cCi$ which is independent of any particular choice of a Cauchy temporal function and its associated GBS splitting. Any other physicist $\bB$, describing the same dynamical phenomenon, accesses the same spatiotemporal object $[\upsilon]$, but he/she does so by means of a GBS splitting associated to a different Cauchy temporal function $\T_2$. More concretely, $\bB$ applies to $[\upsilon]$ the inverse map $[\, . \,]^{-1}_2$, which ``parametrizes'' $[\upsilon]$ so that it becomes $\upsilon_2 \in \Pf(\I_{\T_2})$. The existence of the latter, by the implication $ii) \Rightarrow i)$ in Theorem \ref{mainvar}, assures that $\bB$'s map $\tau \mapsto \nu_\tau$ must be causal as well.

Summarizing the above discussion, we are allowed to say that the notion of a causal measure-valued map $t \mapsto \mu_t$ such that $\supp \, \mu_t \subseteq \T^{-1}(t)$, is a correct way of modelling the dynamics of a spatially distributed physical quantity. Even though it seems tightly associated with a concrete choice of the GBS splitting, it is mathematically equivalent to a splitting-independent spatiotemporal object --- the measure $[\upsilon] \in \Pf(\cCi)$.

\section{Polish spaces of causal curves}
\label{sec::PSoCC}

In the following, every causal curve is implicitly assumed future-directed, $h$ will always denote an auxiliary complete Riemannian metric on a given spacetime $\M$ and $d$ will denote its associated distance function. For any compact $\K \subseteq \M$ and any $r > 0$ define the \emph{generalized $r$-ball centered at $\K$} via $B(\K,r) := \bigcup_{p \in \K} B(p,r)$, where $B(p,r)$ is an ordinary $r$-ball centered at $p \in \M$.

For any interval $I \subseteq \sR$, recall that the compact-open topology on $C(I,\M)$ is nothing but the topology of uniform convergence on compact sets. Without loss of generality, one might consider only the compact subintervals $K = [a,b] \subseteq I$ when checking convergence.

The sets of (possibly not all) causal curves can be topologized in various ways. A seemingly natural way is to endow the set of all causal curves with a fixed domain $I \subseteq \sR$ with the compact-open topology induced from $C(I,\M)$. However, thus obtained space is ``too big'', because various parametrizations of the same causal path are regarded as distinct elements. This problem can be solved by suitably choosing a unique parametrization of each causal path. The standard choice is the arc-length parametrization with respect to an auxiliary Riemannian metric \cite{YCB67,SanchezProgress} and it can even encompass curves with \emph{different} domains \cite{MinguzziCurves08}. Unfortunately, this particular choice of parametrization has a serious drawback. Namely, the limit of a sequence of arc-length-parametrized curves is usually not parametrized by its arc-length. Therefore, this space is not closed in $C(I,\M)$ and hence it is not Polish.

Another standard approach relies on using the so-called $C^0$-topology. In the context of mathematical relativity, it is usually introduced on the space $C(p,q)$ of causal paths with fixed endpoints $p,q \in \M$, where $\M$ is assumed causal or strongly causal \cite{Beem,Geroch1970,HELargeScaleStructure}. Here, however, we follow the exposition from \cite{Penrose1972}, where the endpoints are not fixed.
\begin{Def}
Let $\M$ be a strongly causal spacetime and let $\cC$ denote the set of all compact causal paths. The $C^0$-topology on $\cC$ is defined via its base, which consists of the sets $\cC_U(P,Q)$ of all compact causal paths contained in $U$, with past endpoint in $P$ and future endpoint in $Q$, where $P,Q,U$ are open subsets of $\M$.
\end{Def}
\begin{Rem}
\label{C0Rem1}
The space $C(p,q)$ mentioned above is nothing but $\cC_\M(\{p\},\{q\})$ with the $C^0$-topology induced from $\cC$. More generally, one can endow the spaces $\cC_U(P,Q)$ with $P,Q,U$ \emph{any} subsets of $\M$ with the $C^0$-topology induced from $\cC$.
\end{Rem}
\begin{Rem}
\label{C0Rem2}
Observe that the sequence $([\gamma_n]) \subseteq \cC$ converges to $[\gamma] \in \cC$ iff simultaneuously
\begin{itemize}
\item The sequence of past endpoints of $[\gamma_n]$'s converges in $\M$ to the past endpoint of $[\gamma]$.
\item The sequence of future endpoints of $[\gamma_n]$'s converges in $\M$ to the future endpoint of~$[\gamma]$.
\item For any open $U \supseteq [\gamma]$ it is true that $[\gamma_n] \subseteq U$ for sufficiently large $n$.
\end{itemize}
It is common to define the $C^0$-convergence of \emph{curves} via the $C^0$-convergence of their images. Drawing from the above observation, we say (cf. \cite[Definition 3.33]{Beem}) that the sequence $(\gamma_n) \subseteq C([a,b],\M)$ \emph{converges to $\gamma \in C([a,b],\M)$ in the $C^0$-topology} if
\begin{enumerate}[\itshape i)]
\item $\gamma_n(a) \rightarrow \gamma(a)$ and $\gamma_n(b) \rightarrow \gamma(b)$,
\item for any open $U \supseteq \gamma([a,b])$ it is true that $\gamma_n([a,b]) \subseteq U$ for sufficiently large $n$.
\end{enumerate}
\end{Rem}

\begin{Prop}
\label{C0Prop}
$\cC$ is first-countable.
\end{Prop}
\begin{Proof}\textbf{.}
Fix any $[\gamma] \in \cC$ with endpoints $p,q$. We claim that the sets
\begin{align*}
\cC_{B([\gamma],1/n)}\left(B(p,1/n), B(q,1/n)\right), \ n \in \sN
\end{align*}
constitute the (countable) neighbourhood basis of $[\gamma]$.

Firstly, let $P, Q \subseteq \M$ be any open neighbourhoods of $p,q$, respectively. Clearly, $B(p,1/n) \subseteq P$ and $B(q,1/n) \subseteq Q$ for sufficiently large $n$.

Secondly, let $U \subseteq \M$ be any open set containing $[\gamma]$. We show that $B([\gamma],1/n) \subseteq U$ for $n$ large enough.

By contradiction, assume that there exists an increasing, infinite sequence $(n_k) \subseteq \sN$ such that $B([\gamma],1/n_k) \not\subseteq U$. One can thus construct a sequence $(x_k) \subseteq \M \setminus U$ such that $x_k \in B([\gamma],1/n_k)$ for all $k$. Notice that $(x_k)$ is contained in a precompact set $B([\gamma],1)$ and thus has a subsequence convergent to some $x_{\infty} \in \M \setminus U$, because the latter set is closed. At the same time, since $x_k \in B([\gamma],1/n_k)$ for all $k$, we obtain $x_\infty \in [\gamma] \subseteq U$, which is absurd.

Altogether, we have that for any $P,Q,U \subseteq \M$ open and such that $\cC_U(P,Q) \ni [\gamma]$, for sufficiently large $n$ it is true that $\cC_{B([\gamma],1/n)}(B(p,1/n), B(q,1/n)) \subseteq \cC_U(P,Q)$.
\end{Proof}

It was already Geroch who observed that $C(p,q)$ is separable and metrizable \cite{Geroch1970} (see also \cite{Samann16}). However, to the author's best knowledge, the question whether $\cC$ is Polish has not been addressed. In the following, we show that this is indeed the case (at least in stably causal spacetimes). To this end, we introduce the following new approach to the topologization of the space of causal curves, which employs a \emph{time function} to ``canonically'' parametrize causal paths.
\begin{Def}
\label{CITDef}
Let $\M$ be a stably causal spacetime and let $\T: \M \rightarrow \sR$ be a time function. Fix an interval $I \subseteq \sR$ and define $C^I_\T$ as the space of all causal curves $\gamma: I \rightarrow \M$, along which $\T$ increases at a constant pace, i.e.
\begin{align}
\label{Tcompatible}
\exists c_\gamma > 0 \ \, \forall s,t \in I \ \quad \T(\gamma(t)) - \T(\gamma(s)) = c_\gamma (t-s),
\end{align}
endowed with the compact-open topology induced from $C(I,\M)$.

In the case when $I = [a,b]$, we additionally introduce for any $P,Q \subseteq \M$ the subspace $C^{[a,b]}_\T(P,Q) := C^{[a,b]}_\T \cap \ev_a^{-1}(P) \cap \ev_b^{-1}(Q)$.
\end{Def}
\begin{Rem}
Observe that the constant $c_\gamma$ in (\ref{Tcompatible}) can be expressed as
\begin{align}
\label{cvalue}
c_{\gamma} = \frac{\T(\gamma(b)) - \T(\gamma(a))}{b - a},
\end{align}
\noindent
where $a,b \in I$, $a \neq b$.

This observation leads to an alternative, equivalent formulation of condition (\ref{Tcompatible}), which is sometimes more convenient. Namely,
\begin{align}
\label{alt}
\forall \, t,a,b \in I, \, a \neq b \ \quad \T(\gamma(t)) = \frac{b-t}{b-a}\T(\gamma(a)) + \frac{t-a}{b-a}\T(\gamma(b)).
\end{align}
\end{Rem}

It is not difficult to notice that $\cC$ is in bijection with $C^{[a,b]}_\T$ for any $a,b \in \sR$ and any time function $\T$. Indeed, the proof amounts to showing that the map $[ \, . \, ]: C^{[a,b]}_\T \rightarrow \cC$, $\gamma \mapsto [\gamma]$ is injective. This is a direct consequence of the following proposition.
\begin{Prop}
\label{CITallpaths}
Let $\M$ be a stably causal spacetime, let $\T: \M \rightarrow \sR$ be a time function and fix $a,b \in \sR$. Then for any causal curve $\gamma: [a',b'] \rightarrow \M$ there exists a unique map $\lambda: [a,b] \rightarrow [a',b']$ continuous, strictly increasing and such that $\gamma \circ \lambda \in C^{[a,b]}_\T$.
\end{Prop}
\begin{Proof}\textbf{.}
Because $\T$ is a time function, the map $\T \circ \gamma: [a',b'] \rightarrow [\T(\gamma(a')), \T(\gamma(b'))]$ is continuous, onto and strictly increasing. Hence $(\T \circ \gamma)^{-1}$ exists and is continuous and strictly increasing as well. One can easily convince oneself that in order for property (\ref{alt}) to be satisfied for a curve $\gamma \circ \lambda: [a,b] \rightarrow \M$, it is necessary and sufficient to define $\lambda$ via
\begin{align*}
\forall \, t \in [a,b] \quad \lambda(t) := (\T \circ \gamma)^{-1}\left( \T(\gamma(a')) + \frac{t-a}{b-a} \left(\T(\gamma(b')) - \T(\gamma(a'))\right) \right),
\end{align*}
\noindent
which clearly is also continuous and strictly increasing and hence constitutes a well-defined reparametrization map.
\end{Proof}
\begin{Cor}
\label{CITallpathsCor}
For any $\gamma_1, \gamma_2 \in C^{[a,b]}_\T$ if $[\gamma_1] = [\gamma_2]$ then $\gamma_1 = \gamma_2$.
\end{Cor}
\begin{Proof}\textbf{.}
The equality $[\gamma_1] = [\gamma_2]$ means that there exists a (continuous and strictly increasing) reparametrization map $\lambda: [a,b] \rightarrow [a,b]$ such that $\gamma_1 = \gamma_2 \circ \lambda$. Observe then that $\gamma_2 \circ \lambda \in C^{[a,b]}_\T$, but also, trivially, $\gamma_2 \circ \textrm{id}_{[a,b]} \in C^{[a,b]}_\T$. By Proposition \ref{CITallpaths} we obtain $\lambda = \textrm{id}_{[a,b]}$ and so $\gamma_1 = \gamma_2$.
\end{Proof}

What is less straightforward is that $\cC$ is actually \emph{homeomorphic} to $C^{[a,b]}_\T$. Since the former space is first-countable (Proposition \ref{C0Prop}) and the latter space is metrizable, therefore their topologies are fully determined by the convergent sequences. Their homeomorphicity thus results from the following theorem.
\begin{Thm}
\label{CIThomeo}
Let $\M$ be a stably causal spacetime, let $\T: \M \rightarrow \sR$ be a time function and fix $a,b \in \sR$. Finally, let $(\gamma_n) \subseteq C^{[a,b]}_\T$ and $\gamma \in C^{[a,b]}_\T$. Then $\gamma_n \rightarrow \gamma$ in $C^{[a,b]}_\T$ (i.e. uniformly) iff $[\gamma_n] \rightarrow [\gamma]$ in the $C^0$-topology.
\end{Thm}
\begin{Proof}\textbf{.}
$(\Rightarrow)$
It is enough to show that conditions $i), ii)$ from Remark \ref{C0Rem2} are satisfied. Because uniform convergence implies pointwise convergence, condition $i)$ follows trivially. In order to show $ii)$, take any open $U \supseteq [\gamma]$. From the proof of Proposition \ref{C0Prop}, we already know that $B([\gamma],1/k) \subseteq U$ for $k$ sufficiently large.

Fix such $k$ and notice now that, because $\gamma_n \rightarrow \gamma$ uniformly, then $[\gamma_n] \subseteq B([\gamma],1/k)$ for sufficiently large $n$. Indeed, the former condition means that $\forall t \in [a,b] \ \ d(\gamma_n(t),\gamma(t)) < 1/k$ for sufficiently large $n$, whereas the latter condition is slightly weaker, saying that $\forall t \in [a,b] \ \exists s \in [a,b] \ \ d(\gamma_n(t),\gamma(s)) < 1/k$ for sufficiently large $n$.

Altogether, we have thus obtained that $[\gamma_n] \subseteq U$ for $n$ sufficiently large.

$(\Leftarrow)$ As the first step, we need the following technical result:
\begin{align}
\label{CIThomeo1}
& \forall \, \eps > 0 \ \ \exists \, \eps^\prime, \eps^{\prime\prime} > 0 \ \ \forall \, t \in [a,b]
\\
\nonumber
&\qquad \quad \T^{-1}\left( (\T(\gamma(t)) - \eps^\prime, \T(\gamma(t)) + \eps^\prime ) \right) \, \cap \,  B([\gamma], \eps^{\prime\prime}) \, \subseteq \, B(\gamma(t), \eps).
\end{align}
Suppose (\ref{CIThomeo1}) is not true, i.e. that one can fix $\eps > 0$ such that for any $\eps^\prime, \eps^{\prime\prime} > 0$ there exists $t \in [a,b]$ for which the above inclusion does not hold. Let us fix $\eps^\prime := c_\gamma \delta$, where $c_\gamma$ is given by (\ref{cvalue}) and $\delta > 0$ is such that
\begin{align}
\label{CIThomeo2}
\forall \, \tau_1, \tau_2 \in [a,b] \quad |\tau_1-\tau_2| < \delta \ \Rightarrow \ d(\gamma(\tau_1), \gamma(\tau_2)) < \eps.
\end{align}
\noindent
Notice that the existence of such $\delta$ results directly from the uniform continuity of $\gamma$ (guaranteed, in turn, by the Heine--Borel theorem). For any $k \in \sN$ we also take $\eps^{\prime\prime} := 1/k$ and thus create a sequence $(t_k) \subseteq [a,b]$ such that
\begin{align*}
\T^{-1}\left( (\T(\gamma(t_k)) - c_\gamma \delta, \T(\gamma(t_k)) + c_\gamma \delta ) \right) \, \cap \, B([\gamma], 1/k) \, \not\subseteq \, B(\gamma(t_k), \eps)
\end{align*}
\noindent
for every $k$. This, in turn, means that one can construct a sequence $(p_k) \subseteq B([\gamma], 1)$ such that
\begin{align}
\label{CIThomeo3}
\left| \T(p_k) - \T(\gamma(t_k)) \right| < c_\gamma \delta \quad \wedge \quad p_k \in B([\gamma], 1/k) \quad \wedge \quad d(p_k, \gamma(t_k)) \geq \eps
\end{align}
\noindent
for every $k$. Because $[a,b]$ is compact and $B([\gamma], 1)$ is precompact, we can pass to subsequences of $(t_k)$ and $(p_k)$ that (simultaneously) converge to some $t_\infty \in [a,b]$ and $p_\infty \in \overline{B([\gamma], 1)}$, respectively. From (\ref{CIThomeo3}) we obtain that in fact $p_\infty \in [\gamma]$ and so $p_\infty = \gamma(s)$ for some $s \in [a,b]$. Furthermore, (\ref{CIThomeo3}) yields also that
\begin{align*}
\left| \T(\gamma(s)) - \T(\gamma(t_\infty)) \right| < c_\gamma \delta \quad \wedge \quad d(\gamma(s), \gamma(t_\infty)) \geq \eps.
\end{align*}
\noindent
However, using (\ref{Tcompatible}) we can rewrite the left-hand side of the first condition as $c_\gamma |s - t_\infty|$ and obtain that
\begin{align*}
\left| s - t_\infty \right| < \delta \quad \wedge \quad d(\gamma(s), \gamma(t_\infty)) \geq \eps,
\end{align*}
\noindent
which contradicts (\ref{CIThomeo2}) and thus completes the proof of (\ref{CIThomeo1}).

We are now ready to prove that $\gamma_n \rightarrow \gamma$ uniformly, i.e. that
\begin{align}
\label{CIThomeo4}
\forall \, \eps>0 \ \exists \, N \in \sN \ \forall \, n \geq N \ \forall \, t \in [a,b] \quad d(\gamma(t),\gamma_n(t)) < \eps.
\end{align}
To this end, firstly observe that condition $i)$ from Remark \ref{C0Rem2} implies that $\T \circ \gamma_n \rightarrow \T \circ \gamma$ uniformly, i.e.
\begin{align}
\label{CIThomeo5}
\forall \, \eps^\prime>0 \ \exists \, N^\prime \in \sN \ \forall \, n \geq N^\prime \ \forall \, t \in [a,b] \quad \left| \T(\gamma(t)) - \T(\gamma_n(t)) \right| < \eps^\prime.
\end{align}
Indeed, employing (\ref{alt}), we easily get that for any $t \in [a,b]$
\begin{align*}
\left| \T(\gamma(t)) - \T(\gamma_n(t)) \right| \leq \left| \T(\gamma(a)) - \T(\gamma_n(a)) \right| + \left| \T(\gamma(b)) - \T(\gamma_n(b)) \right|,
\end{align*}
\noindent
what on the strength of $i)$ already yields the uniform convergence.

Additionally, observe that condition $ii)$ from Remark \ref{C0Rem2} implies that
\begin{align}
\label{CIThomeo6}
\forall \, \eps^{\prime \prime}>0 \ \exists \, N^{\prime\prime} \in \sN \ \forall \, n \geq N^{\prime\prime} \quad [\gamma_n] \subseteq B([\gamma],\eps^{\prime \prime}).
\end{align}
Indeed, simply take $U := B([\gamma],\eps^{\prime \prime})$.

To prove (\ref{CIThomeo4}), fix any $\eps>0$, take $\eps^\prime, \eps^{\prime\prime} > 0$ as in (\ref{CIThomeo1}), then take $N^\prime, N^{\prime\prime} \in \sN$ as in (\ref{CIThomeo5},\ref{CIThomeo6}) and define $N:=\max\{N^\prime,N^{\prime \prime}\}$. For any $n \geq N$, (\ref{CIThomeo5},\ref{CIThomeo6}) imply that for all $t \in [a,b]$
\begin{align*}
\gamma_n(t) \, \in \, \T^{-1}\left( (\T(\gamma(t)) - \eps^\prime, \T(\gamma(t)) + \eps^\prime ) \right) \, \cap \,  B([\gamma], \eps^{\prime\prime}),
\end{align*}
which, on the strength of (\ref{CIThomeo1}), gives that $\gamma_n(t) \in B(\gamma(t), \eps)$ for all $t \in [a,b]$, completing the proof of (\ref{CIThomeo4}).
\end{Proof}

Two straightforward corollaries follow.
\begin{Cor}
\label{CIThomeoCor1}
Let $\M, \T, a, b$ be as above and let $P,Q$ be \emph{any} subsets of $\M$. Then $\cC_\M(P,Q) \cong C_\T^{[a,b]}(P,Q)$.
\end{Cor}
\begin{Proof}\textbf{.}
Because we are only putting constraints on the location of endpoints, the restricted map $[\, . \,]: C_\T^{[a,b]}(P,Q) \rightarrow \cC_\M(P,Q)$ is still a well-defined bijection. Moreover, elementary properties of the subspace topology guarantee that it is still a homeomorphism.
\end{Proof}
\begin{Cor}
\label{CIThomeoCor2}
Let $P,Q$ be subsets of a stably causal spacetime $\M$, let $\T_1, \T_2: \M \rightarrow \sR$ be time functions and fix $a_i,b_i \in \sR$, $a_i < b_i$ for $i=1,2$. Then the spaces $C^{[a_1,b_1]}_{\T_1}(P,Q)$ and $C^{[a_2,b_2]}_{\T_2}(P,Q)$ are homeomorphic.
\end{Cor}
\begin{Proof}\textbf{.}
By the previous corollary, $C^{[a_1,b_1]}_{\T_1}(P,Q) \cong \cC_\M(P,Q) \cong C^{[a_2,b_2]}_{\T_2}(P,Q)$.
\end{Proof}

Furthermore, Corollary \ref{CIThomeoCor1} allows to restate Leray's characterization of global hyperbolicity in terms of the spaces $C^{[a,b]}_\T$.
\begin{Prop}
\label{Leray1}
Let $\M$ be a stably causal spacetime and let $\T: \M \rightarrow \sR$ be a time function. Then the following conditions are equivalent.
\begin{enumerate}[\itshape i)]
\item $\M$ is globally hyperbolic.
\item $C^{[a,b]}_\T(K_1,K_2)$ is compact for all $a,b \in \sR$ and all compact $K_1,K_2 \subseteq \M$.
\item $C^{[0,1]}_\T(p,q)$ is compact for all $p,q \in \M$.
\end{enumerate}
\end{Prop}
\begin{Proof}\textbf{.}
With Corollary \ref{CIThomeoCor1} guaranteeing that $C^{[0,1]}_\T(p,q) \cong C(p,q)$, the equivalence $i) \Leftrightarrow iii)$ follows from \cite{Geroch1970} (compare also \cite[Proposition~6.6.2]{HELargeScaleStructure} and \cite[Theorem~3.79]{MS08}). The implication $i) \Rightarrow ii)$ can be proven by first invoking Corollary \ref{CIThomeoCor1} to get $C^{[a,b]}_\T(K_1,K_2) \cong \cC_\M(K_1,K_2) = \cC_{J^+(K_1) \cap J^-(K_2)}(K_1,K_2)$ and then noticing that the latter space is compact on the strength of \cite[Theorem~6.5]{Penrose1972} and Proposition \ref{PropCCC} (see \hyperref[sec::preliminaries]{Appendix}). Finally, the implication $ii) \Rightarrow iii)$ is trivial.
\end{Proof}

We now address the question of the Polishness of $C^I_\T$.
\begin{Prop}
\label{PropCITPolish}
Let $\M$ be a stably causal spacetime, let $\T: \M \rightarrow \sR$ be a time function and let $I \subseteq \sR$ be an interval. Then $C^I_\T$ is a Polish space. Moreover, if $I = [a,b]$ then the space $C^{[a,b]}_\T(P,Q)$ is Polish for any \emph{closed} $P,Q \subseteq \M$.
\end{Prop}
\begin{Proof}\textbf{.}
We show below that $C^I_\T$ is a closed subspace of the Polish space $C(I,\M)$. The Polishness of $C^{[a,b]}_\T(P,Q)$ for any closed $P,Q \subseteq \M$ will then follow from the continuity of the evaluation maps $\textnormal{ev}_a$, $\textnormal{ev}_b$.

To this end, suppose that $(\gamma_n) \subseteq C^I_\T$ converges to $\gamma \in C(I,\M)$ in the compact-open topology. In order to show that $\gamma$ satisfies (\ref{alt}), fix any $t,a,b \in I, a \neq b$ and use the fact that the uniform convergence on compact sets implies the pointwise convergence to obtain
\begin{align*}
\T(\gamma(t)) & = \lim\limits_{n \rightarrow +\infty} \T(\gamma_n(t)) = \lim\limits_{n \rightarrow +\infty} \frac{b-t}{b-a}\T(\gamma_n(a)) + \frac{t-a}{b-a}\T(\gamma_n(b))
\\
& = \frac{b-t}{b-a}\T(\gamma(a)) + \frac{t-a}{b-a}\T(\gamma(b)).
\end{align*}

We still have to prove that $\gamma$ is future-directed and causal. To achieve this, we will use Proposition \ref{futdircaus} (see \hyperref[sec::preliminaries]{Appendix}).

Let us define two sequences $(a_m), (b_m) \subseteq I$, $a_m < b_m$ for all $m \in \sN$ as follows. If $\inf I \in I$, define $a_m \equiv \inf I$, otherwise let $(a_m)$ be a decreasing sequence tending to $\inf I$. Similarly, if $\sup I \in I$, define $b_m \equiv \sup I$, otherwise let $(b_m)$ be an increasing sequence tending to $\sup I$.

By assumption, for every $m \in \sN$ the sequence $\left( \gamma_n|_{[a_m,b_m]} \right)_{n \in \sN}$ converges to $\gamma|_{[a_m,b_m]}$ uniformly, and hence also in the $C^0$-topology by Theorem \ref{CIThomeo}. On the strength of \cite[Proposition 3.34 \& Lemma 3.29]{Beem}, the latter convergence implies that $\gamma|_{[a_m,b_m]}$ is future-directed causal for every $m \in \sN$. This, in turn, clearly shows that $\gamma$ satisfies the characterization of a causal curve (\hyperref[sec::preliminaries]{Appendix}, Proposition \ref{futdircaus}), because any $s,t \in I$ lie in $[a_m,b_m]$ for $m$ large enough.
\end{Proof}

As an immediate corollary of Theorem \ref{CIThomeo} (with Corollary \ref{CIThomeoCor2}) and Proposition \ref{PropCITPolish} we obtain:
\begin{Cor}
Let $\M$ be a stably causal spacetime. Then for any closed $P,Q \subseteq \M$ the space $\cC_\M(P,Q)$ is Polish. In particular, $\cC$ is a Polish space.
\end{Cor}

We see that the spaces $C_\T^I$ with $I = [a,b]$ cast some new light on $\cC$, i.e. the space of all \emph{compact} causal paths. But what about \emph{noncompact} causal paths?

Below we provide a ``noncompact'' analogue (or rather a complement) of Proposition \ref{CITallpaths}, which answers the question whether a \emph{noncompact} causal path can be uniquely parametrized so as to become an element of $C^I_\T$ for a suitable interval $I$. The answer turns out to be rather subtle, heavily depending on the time function $\T$ employed. Namely, if $\T$ is \emph{bounded} on a given causal path, such a unique parametrization exists. If, on the other hand, $\T$ is unbounded on a causal path, the parametrization of the latter is unique only up to a certain affine transformation.
\begin{Prop}
\label{CITallpaths3}
Let $\M$ be a stably causal spacetime and let $\T: \M \rightarrow \sR$ be a time function. Let $I':=(a',b')$, $-\infty \leq a' < b' \leq +\infty$ and let $\gamma: I' \rightarrow \M$ be a causal curve. The map $\T \circ \gamma$ is continuous, strictly increasing and hence onto $(T_{\gamma}, T^{\gamma})$, where $T_{\gamma} := \lim_{\tau \searrow a'} \T(\gamma(\tau))$ and $T^{\gamma} := \lim_{\tau \nearrow b'} \T(\gamma(\tau))$ might be equal to $-\infty$ and $+\infty$, respectively. One has that:
\begin{itemize}
\item If both $T_{\gamma}$ and $T^{\gamma}$ are finite, then $\gamma \circ \lambda \in C^I_\T$ if and only if $I = (a,b)$, $a,b \in \sR$ and $\lambda: I \rightarrow I'$ is of the form
    \begin{align*}
    \forall \, t \in I \quad \lambda(t) := (\T \circ \gamma)^{-1}\left( T_{\gamma} + \frac{t-a}{b-a} \left(T^{\gamma} - T_{\gamma}\right) \right).
    \end{align*}
\noindent
The same is true if $I' = [a',b')$ ($I' = (a',b']$), only now $I = [a,b)$ ($I = (a,b]$).
\item If $T_{\gamma}$ is finite and $T^{\gamma} = +\infty$, then $\gamma \circ \lambda \in C^I_\T$ if and only if $I = (a,+\infty)$, $a \in \sR$ and $\lambda: I \rightarrow I'$ is of the form
    \begin{align*}
    \forall \, t \in I \quad \lambda(t) := (\T \circ \gamma)^{-1}\left( T_{\gamma} + A(t - a) \right),
    \end{align*}
\noindent
where $A$ is an arbitrary positive constant. The same is true if $I' = [a',b')$, only now $I = [a,+\infty)$.
\item If $T_{\gamma} = -\infty$ and $T^{\gamma}$ is finite, then $\gamma \circ \lambda \in C^I_\T$ if and only if $I = (-\infty,b)$, $b \in \sR$ and $\lambda: I \rightarrow I'$ is of the form
    \begin{align*}
    \forall \, t \in I \quad \lambda(t) := (\T \circ \gamma)^{-1}\left( T^{\gamma} + A(t - b) \right),
    \end{align*}
\noindent
where $A$ is an arbitrary positive constant. The same is true if $I' = (a',b']$, only now $I = (-\infty,b]$.
\item If $T_{\gamma} = -\infty$ and $T^{\gamma} = +\infty$, then $\gamma \circ \lambda \in C^I_\T$ if and only if $I = \sR$ and $\lambda: \sR \rightarrow I'$ is of the form
    \begin{align*}
    \forall \, t \in \sR \quad \lambda(t) := (\T \circ \gamma)^{-1}\left( A t + B \right),
    \end{align*}
\noindent
where $A,B$ are arbitrary real constants with $A > 0$.
\end{itemize}
\end{Prop}
\begin{Proof}\textbf{.}
That $\T \circ \gamma$ is continuous and strictly increasing results from the very definition of a time function. Hence, the inverse map $(\T \circ \gamma)^{-1}: (T_{\gamma}, T^{\gamma}) \rightarrow I'$ exists, it is continuous and strictly increasing.

Similarly as in the proof of Proposition \ref{CITallpaths}, one can easily convince oneself that in each particular case, the given formula for $\lambda$ is necessary and sufficient for the curve $\gamma \circ \lambda$ to satisfy (\ref{Tcompatible}) under the assumptions pertaining to that case.

Finally, notice that if $I' = [a',b')$, then one has to replace $(T_{\gamma}, T^{\gamma})$ with $[T_{\gamma}, T^{\gamma})$, and also define $I$ as containing its left endpoint. Similar simple modifications are needed in the $I' = (a',b']$ case.
\end{Proof}

As we can see, if the interval $I$ is noncompact, then the question which causal paths are ``included'' in $C_\T^I$ and which are not is rather complicated, depending on the details of the chosen time function $\T$ and the interval $I$. In particular, there is no simple analogue of Corollary \ref{CIThomeoCor2}. Although addressing these questions in full generality lies beyond the scope of this paper, below we answer them assuming some additional properties of $\T$.
\begin{Prop}
\label{CITallpaths4}
Let $\M$ be a globally hyperbolic spacetime and let $\T: \M \rightarrow \sR$ be a \emph{Cauchy} time function. Then the map $C^{\sR}_\T \ni \gamma \mapsto [\gamma]$ is a surjection onto the set of all inextendible causal paths on $\M$.
\end{Prop}
\begin{Proof}\textbf{.}
We need to show two things: that every $\gamma \in C^{\sR}_\T$ is an inextendible causal curve (and hence $[\gamma]$ is an inextendible causal path) and that every inextendible causal path can be parametrized so as to become an element of $C^{\sR}_\T$. In fact, the latter claim is a direct consequence of Proposition \ref{CITallpaths3} (the last bullet), because $\T$, being a Cauchy time function, assumes all real values on every inextendible causal path. As a side remark, notice that the arbitrariness of the constants $A,B$ in Proposition \ref{CITallpaths3} means that one cannot replace `surjection' with `bijection' in the above statement.

As for the first claim, suppose that $\gamma \in C^{\sR}_\T$ has a future endpoint $q := \lim_{t \rightarrow +\infty} \gamma(t)$. By (\ref{Tcompatible}), we can write that
\begin{align*}
\forall \, t \in \sR \quad \T(\gamma(t)) = \T(\gamma(0)) + c_{\gamma}t.
\end{align*}
But passing now with $t$ to $+\infty$ and using the continuity of $\T$, we obtain
\begin{align*}
\T(q) = \lim\limits_{t \rightarrow +\infty} \T(\gamma(t)) =  \lim\limits_{t \rightarrow +\infty} \left[ \T(\gamma(0)) + c_{\gamma} t \right] = +\infty,
\end{align*}
a contradiction.

Similarly one can show that $\gamma$ does not have a past endpoint.
\end{Proof}

\begin{Prop}
\label{CITallpaths5}
Let $\M$ be a globally hyperbolic spacetime and let $\T_1, \T_2 : \M \rightarrow \sR$ be Cauchy time functions. Then the map $\widetilde{\ }: C^{\sR}_{\T_1} \rightarrow C^{\sR}_{\T_2}$, defined via
\begin{align}
\label{Rhomeo}
\forall \, \gamma \in C^{\sR}_{\T_1} \quad \widetilde{\gamma} := \gamma \circ \left(\T_2 \circ \gamma \right)^{-1} \circ \T_1 \circ \gamma,
\end{align}
\noindent
is a bijection.
\end{Prop}
\begin{Proof}\textbf{.}
First, observe that formula (\ref{Rhomeo}) produces well-defined elements of $C^{\sR}_{\T_2}$. Indeed, since $\T_1, \T_2$ are Cauchy time functions and any $\gamma \in C^{\sR}_{\T_1}$ is inextendible (Proposition \ref{CITallpaths4}), the maps $(\T_i \circ \gamma)$, $i=1,2$ are continuous strictly increasing bijections of $\sR$ onto itself and so are their inverses. Hence $\widetilde{\gamma}$ is a reparametrization of $\gamma$ and thus it is an inextendible causal curve. Moreover, observe that $\T_2 \circ \widetilde{\gamma} = \T_1 \circ \gamma$ and therefore
\begin{align*}
\forall \, s,t \in \sR \quad \T_2(\widetilde{\gamma}(t)) - \T_2(\widetilde{\gamma}(s)) = \T_1(\gamma(t)) - \T_1(\gamma(s)) = c_\gamma (t-s),
\end{align*}
which completes the proof that $\widetilde{\gamma} \in C^{\sR}_{\T_2}$.

Having shown that $\gamma \mapsto \widetilde{\gamma}$ is a well-defined map, we immediately see that it is a bijection, its inverse being the map
\begin{align*}
C^{\sR}_{\T_2} \ni \rho \mapsto \rho \circ \left(\T_1 \circ \rho \right)^{-1} \circ \T_2 \circ \rho.
\end{align*}
Notice that (quite expectedly) the formulas for $\widetilde{\ }$ and its inverse differ only in that $\T_1$ and $\T_2$ are swapped, and so the inverse map is well defined by a reasoning completely analogous to the one conducted above.
\end{Proof}
Let us remark that (\ref{Rhomeo}) might also be written in the form $\widetilde{\gamma} := \T_2|_{[\gamma]}^{-1} \circ \T_1|_{[\gamma]} \circ \gamma$.

Our goal is to show that $\widetilde{\ }$ is actually a \emph{homeomorphism}, provided $\T_1, \T_2$ are Cauchy \emph{temporal} functions. To this end we shall need several technical lemmas, some of which, however, elucidate some further properties of the spaces $C^{I}_{\T}$.

\begin{Lem}
\label{RhomeoLem2}
Let $\M$ be a stably causal spacetime, $\T: \M \rightarrow \sR$ a time function and $I \subseteq \sR$ an interval. Assume that $\gamma_n \rightarrow \gamma$ in $C_\T^I$. Then the positive sequence $(c_{\gamma_n})$ converges to $c_\gamma > 0$. 
\end{Lem}
\begin{Proof}\textbf{.}
Fix any $a,b \in I$. Using (\ref{cvalue}), the continuity of $\T$ and the fact that convergence in the compact-open topology implies pointwise convergence, one obtains
\begin{align*}
c_{\gamma_n} = \frac{\T(\gamma_n(b)) - \T(\gamma_n(a))}{b - a} \rightarrow \frac{\T(\gamma(b)) - \T(\gamma(a))}{b - a} = c_\gamma.
\end{align*}
\end{Proof}

In the remaining lemmas, $\M$ will always be a globally hyperbolic spacetime.
\begin{Lem}
\label{RhomeoLem2iPol}
Let $\T_1, \T_2: \M \rightarrow \sR$ be smooth time functions and let $I \subseteq \sR$ be an interval. Assume that $\gamma_n \rightarrow \gamma$ in $C_{\T_1}^I$. Then $\T_2 \circ \gamma_n \rightarrow \T_2 \circ \gamma$ in $C(I,\sR)$ uniformly on compact sets.
\end{Lem}
\begin{Proof}\textbf{.}
It suffices to prove that $\T_2 \circ \gamma_n \rightarrow \T_2 \circ \gamma$ uniformly on any $[a,b] \subseteq I$. To begin with, observe that, since $\gamma_n|_{[a,b]} \rightarrow \gamma|_{[a,b]}$ uniformly, therefore the sequences $(\gamma_n(a))$, $(\gamma_n(b)) \subseteq \M$ are convergent and hence bounded. Define
\begin{align}
\label{RL2iPolK}
\K & := J^+\left(\overline{\{\gamma_n(a) \, | \, n \in \sN\}}\right) \cap J^-\left(\overline{\{\gamma_n(b) \, | \, n \in \sN\}}\right),
\end{align}
which is a compact subset of $\M$ on the strength of Proposition \ref{PropCCC}, \emph{i)} (see \hyperref[sec::preliminaries]{Appendix}). What is more, $\K$ contains the images $\gamma([a,b])$ and $\gamma_n([a,b])$ for all $n \in \sN$. Knowing that $\T_2$, being smooth, is locally Lipschitz continuous, we have
\begin{align}
\label{locLip}
\exists \, L > 0 \ \forall \, p,q \in \K \quad \left| \T_2(p) - \T_2(q) \right| \leq L \, d(p,q),
\end{align}
and so
\begin{align*}
\sup\limits_{t \in [a,b]} \left| \T_2(\gamma_n(t)) - \T_2(\gamma(t)) \right| \leq L \sup\limits_{t \in [a,b]} d(\gamma_n(t),\gamma(t)) \rightarrow 0.
\end{align*}
\end{Proof}

\begin{Lem}
\label{RhomeoLem3}
Let $\T: \M \rightarrow \sR$ be a Cauchy temporal function and let $I \subseteq \sR$ be an interval. Assume that $\gamma_n \rightarrow \gamma$ in $C_\T^I$. Then
\begin{align}
\label{RL3}
\forall \, [a,b] \subseteq I \ \exists l_{a,b} > 0 \ \forall \, n \in \sN \ \forall s,t \in [a,b] \quad \frac{1}{l_{a,b}} |s - t| \leq d_w(\gamma_n(s),\gamma_n(t)) \leq l_{a,b} |s - t|,
\end{align}
where $d_w$ is the distance function associated to the complete Riemannian metric $w$ given by (\ref{Rmetric2}).
\end{Lem}
\begin{Proof}\textbf{.}
Fix $[a,b] \subseteq I$ and define the compact set $\K$ by (\ref{RL2iPolK}). The first inequality in (\ref{RL3}) can be proven by noticing that for any $s,t \in [a,b]$ and any $n \in \sN$
\begin{align*}
d_w(\gamma_n(s),\gamma_n(t)) \geq \frac{1}{L} \left| \T(\gamma_n(s)) - \T(\gamma_n(t)) \right| = \frac{c_{\gamma_n}}{L} |s - t| \geq \underbrace{\frac{1}{L} \inf\limits_{n \in \sN} c_{\gamma_n}}_{=: \, l_1} \cdot |s - t|,
\end{align*}
where $L$ is a Lipschitz constant for $\T|_\K$ (cf. (\ref{locLip})), $\inf_{n \in \sN} \, c_{\gamma_n} > 0$ by Lemma \ref{RhomeoLem2} and the equality follows from (\ref{Tcompatible}).

We now move to the second inequality in (\ref{RL3}). Fix $n \in \sN$ and recall that $\gamma_n^\prime(\tau)$ exists and is a future-directed causal vector for $\tau \in [a,b]$ a.e. \cite[Remark 3.18]{MS08}, which means that
\begin{align}
\label{RL3a}
g\left(\gamma^\prime_n(\tau), \gamma^\prime_n(\tau) \right) \leq 0
\end{align}
for $\tau \in [a,b]$ a.e. Notice, moreover, that for such $\tau$ (\ref{cvalue}) implies that
\begin{align}
\label{RL3b}
d\T(\gamma^\prime_n(\tau)) = (\T \circ \gamma_n)^\prime(\tau) = \lim\limits_{\xi \rightarrow \tau} \frac{\T(\gamma_n(\xi)) - \T(\gamma_n(\tau))}{\xi - \tau} = c_{\gamma_n}.
\end{align}

Assuming $s < t$ and using (\ref{Rmetric2}), (\ref{RL3a}) and (\ref{RL3b}), one obtains
\begin{align*}
& d_w(\gamma_n(s), \gamma_n(t)) \leq \int_s^t \sqrt{w\left(\gamma^\prime_n(\tau), \gamma^\prime_n(\tau) \right)} d\tau
\\
& = \int_s^t \sqrt{u(\gamma_n(\tau))} \cdot \sqrt{g\left(\gamma^\prime_n(\tau), \gamma^\prime_n(\tau) \right) + 2 \alpha(\gamma_n(\tau)) \left[d\T(\gamma^\prime_n(\tau))\right]^2} d\tau
\\
& \leq c_{\gamma_n} \int_s^t \sqrt{2 u(\gamma_n(\tau)) \cdot \alpha(\gamma_n(\tau))} \, d\tau \leq \underbrace{\sup\limits_{n \in \sN} c_{\gamma_n} \cdot \max\limits_{p \in \K} \sqrt{2u(p)\alpha(p)}}_{=: \, l_2} \cdot | s - t |,
\end{align*}
where the compactness of $\K$ assures that the continuous map $p \mapsto \sqrt{2u(p)\alpha(p)}$ attains its maximum on $\K$. The sequence $(c_{\gamma_n})$ is bounded by Lemma \ref{RhomeoLem2}.

To finish the proof of (\ref{RL3}), we obviously take $l_{a,b} := \max\{ 1/l_1, l_2 \}$.
\end{Proof}

Property (\ref{RL3}) could be called the \emph{local bi-Lipschitz equicontinuity} of $\{\gamma_n\}_{n \in \sN} \subseteq C_\T^I$ (compare \cite[pp. 75--76]{Beem}). The next lemma shows that this property remains valid if we compose $\gamma_n$'s with another Cauchy temporal function.
\begin{Lem}
\label{RhomeoLem4}
Let $\T_1, \T_2: \M \rightarrow \sR$ be Cauchy temporal functions and let $I \subseteq \sR$ be an interval. Assume that $\gamma_n \rightarrow \gamma$ in $C_{\T_1}^I$. Then
\begin{align}
\nonumber
&\forall \, [a,b] \subseteq I \ \exists L_{a,b} > 0 \ \forall \, n \in \sN \ \forall s,t \in [a,b]
\\
\label{RL4}
&\qquad \quad \frac{1}{L_{a,b}} |s - t| \leq \left| \T_2(\gamma_n(s)) - \T_2(\gamma_n(t)) \right| \leq L_{a,b} |s - t|.
\end{align}
\end{Lem}
\begin{Proof}\textbf{.}
Fix $[a,b] \subseteq I$. For any fixed $n \in \sN$, by (\ref{Geroch2}), (\ref{RL3a}) and (\ref{RL3b}) we know that
\begin{align*}
0 & \geq g\left(\gamma^\prime_n(\tau), \gamma^\prime_n(\tau) \right) = -\alpha(\gamma_n(\tau)) \left[ d\T_1(\gamma^\prime_n(\tau)) \right]^2 + \bar{g}\left(\gamma^\prime_n(\tau), \gamma^\prime_n(\tau) \right)
\\
& = -\alpha(\gamma_n(\tau)) c_{\gamma_n}^2 + \bar{g}\left(\gamma^\prime_n(\tau), \gamma^\prime_n(\tau) \right)
\end{align*}
for $\tau \in [a,b]$ a.e. and hence, for such $\tau$,
\begin{align}
\label{RL4a}
\sqrt{\bar{g}\left(\gamma^\prime_n(\tau), \gamma^\prime_n(\tau) \right)} \leq c_{\gamma_n} \sqrt{\alpha(\gamma_n(\tau))}.
\end{align}
Furthermore, (\ref{Geroch2}) gives us that
\begin{align}
\label{RL4b}
[d\T_1 (\grad \, \T_2)]^2 = \frac{\bar{g}(\grad \, \T_2,\grad \, \T_2) - g(\grad \, \T_2,\grad \, \T_2)}{\alpha}.
\end{align}
Notice that $d\T_1 (\grad \, \T_2) = (\grad \, \T_2)(\T_1) < 0$, because $\T_1,\T_2$ are temporal functions. Therefore, taking the square root of (\ref{RL4b}) yields
\begin{align}
\label{RL4c}
d\T_1 (\grad \, \T_2) = - \frac{1}{\sqrt{\alpha}} \sqrt{\bar{g}(\grad \, \T_2,\grad \, \T_2) - g(\grad \, \T_2,\grad \, \T_2)}.
\end{align}
Altogether, for any $\tau \in [a,b]$ at which $\gamma^\prime_n$ exists we obtain that
\begin{align}
\nonumber
& (\T_2 \circ \gamma_n)^\prime(\tau) = g(\grad \, \T_2, \gamma^\prime_n) = -\alpha \, d\T_1 (\grad \, \T_2) \, d\T_1 (\gamma^\prime_n) + \bar{g}(\grad \, \T_2, \gamma^\prime_n)
\\
\label{RL4d}
& \qquad \geq - c_{\gamma_n} \alpha \, d\T_1 (\grad \, \T_2) - \sqrt{\bar{g}(\grad \, \T_2, \grad \, \T_2)} \, \sqrt{\bar{g}(\gamma^\prime_n, \gamma^\prime_n)}
\\
\nonumber
& \qquad \geq c_{\gamma_n} \underbrace{\sqrt{\alpha} \left[ \sqrt{\bar{g}(\grad \, \T_2,\grad \, \T_2) - g(\grad \, \T_2,\grad \, \T_2)} - \sqrt{\bar{g}(\grad \, \T_2, \grad \, \T_2)} \right]}_{=: \, G},
\end{align}
where we have suppressed the arguments of the functions and then used (\ref{Geroch2}), (\ref{RL3b}), the Cauchy--Schwarz inequality for $\bar{g}$, (\ref{RL4a}) and (\ref{RL4c}). Observe that the newly defined function $G: \M \rightarrow \sR$ is continuous and positive, because $\T_2$ is temporal and hence $g(\grad \, \T_2, \grad \, \T_2)$ is negative.

Assuming $s < t$, with the help of (\ref{RL4d}) we can now obtain the first inequality in (\ref{RL4}). Namely,
\begin{align*}
& \T_2(\gamma_n(t)) - \T_2(\gamma_n(s)) = \int_s^t (\T_2 \circ \gamma_n)^\prime(\tau) d\tau \geq \underbrace{\inf\limits_{n \in \sN} c_{\gamma_n} \cdot \min\limits_{r \in \K} G(r)}_{=: \, L_1} \cdot \, (t - s),
\end{align*}
where $\K$ is the compact set defined by (\ref{RL2iPolK}) and thus $G$ attains on $\K$ its minimum, which is positive. Lemma \ref{RhomeoLem2} assures the positivity of $\inf_n c_{\gamma_n}$.

As for the second inequality in (\ref{RL4}), it is a straightforward consequence of (\ref{locLip}) and Lemma \ref{RhomeoLem3}, on the strength of which one has
\begin{align*}
\forall \, n \in \sN \ \forall s,t \in [a,b] \quad \left| \T_2(\gamma_n(s)) - \T_2(\gamma_n(t)) \right| \leq L \, d_w(\gamma_n(s),\gamma_n(t)) \leq \underbrace{L \, l_{A,B}}_{=: \, L_2} |s-t|,
\end{align*}
where $A := \min_{r \in \K} \T_2(r)$ and $B := \max_{r \in \K} \T_2(r)$.

To finish the proof of (\ref{RL4}), we obviously take $L_{a,b} := \max\{ 1/L_1, L_2 \}$.
\end{Proof}

\begin{Lem}
\label{RhomeoLem5}
Under the assumptions of Lemma \ref{RhomeoLem4} with $I := \sR$ it is true that
\begin{align*}
(\T_2 \circ \gamma_n)^{-1} \circ \T_1 \circ \gamma_n \rightarrow (\T_2 \circ \gamma)^{-1} \circ \T_1 \circ \gamma
\end{align*}
in $C(\sR,\sR)$ uniformly on compact sets.
\end{Lem}
\begin{Proof}\textbf{.}
It suffices to show that $(\T_2 \circ \gamma_n)^{-1} \circ \T_1 \circ \gamma_n \rightarrow (\T_2 \circ \gamma)^{-1} \circ \T_1 \circ \gamma$ uniformly on any $[a,b] \subseteq \sR$. For any $t \in [a,b]$ one has
\begin{align}
\nonumber
& \left| (\T_2 \circ \gamma_n)^{-1} (\T_1(\gamma_n(t))) - (\T_2 \circ \gamma)^{-1} (\T_1(\gamma(t))) \right|
\\
\label{RL5a}
& \qquad \quad \leq \left| (\T_2 \circ \gamma_n)^{-1} (\T_1(\gamma_n(t))) - (\T_2 \circ \gamma_n)^{-1} (\T_1(\gamma(t))) \right|
\\
\nonumber
& \qquad \quad + \left| (\T_2 \circ \gamma_n)^{-1} (\T_1(\gamma(t))) - (\T_2 \circ \gamma)^{-1} (\T_1(\gamma(t))) \right|.
\end{align}
Both terms on the right-hand side can be estimated from above with the help of Lemma~\ref{RhomeoLem4} (first inequality). For the first term one has
\begin{align*}
\left| (\T_2 \circ \gamma_n)^{-1} (\T_1(\gamma_n(t))) - (\T_2 \circ \gamma_n)^{-1} (\T_1(\gamma(t))) \right| \leq L_{A,B} \left| \T_1(\gamma_n(t)) - \T_1(\gamma(t)) \right|,
\end{align*}
where $A := \min_{r \in \K} \T_1(r)$ and $B := \max_{r \in \K} \T_1(r)$, in which $\K$ is again the compact set given by (\ref{RL2iPolK}).

To estimate the second term, it is convenient to introduce the auxiliary variable $\tau := (\T_2 \circ \gamma)^{-1} (\T_1(\gamma(t)))$. Observe that $\tau \in [a',b']$, where $a' := (\T_2 \circ \gamma)^{-1} (\T_1(\gamma(a)))$ and $b' := (\T_2 \circ \gamma)^{-1} (\T_1(\gamma(b)))$, because the map $(\T_2 \circ \gamma)^{-1} \circ \T_1 \circ \gamma$ is continuous and strictly increasing. One can now write that
\begin{align*}
& \left| (\T_2 \circ \gamma_n)^{-1} (\T_1(\gamma(t))) - (\T_2 \circ \gamma)^{-1} (\T_1(\gamma(t))) \right| = \left| (\T_2 \circ \gamma_n)^{-1} (\T_2(\gamma(\tau))) - \tau \right|
\\
& = \left| (\T_2 \circ \gamma_n)^{-1} (\T_2(\gamma(\tau))) - (\T_2 \circ \gamma_n)^{-1} (\T_2(\gamma_n(\tau))) \right| \leq L_{A',B'} \left| \T_2(\gamma(\tau)) - \T_2(\gamma_n(\tau)) \right|,
\end{align*}
where $A' := \min_{r \in \K'} \T_2(r)$ and $B' := \max_{r \in \K'} \T_2(r)$, in which $\K'$ is defined as
\begin{align*}
\K' := J^+\left(\overline{\{\gamma_n(a') \, | \, n \in \sN\}}\right) \cap J^-\left(\overline{\{\gamma_n(b') \, | \, n \in \sN\}}\right),
\end{align*}
i.e. it is another compact subset of $\M$ designed so as to contain the images $\gamma([a',b'])$ and $\gamma_n([a',b'])$ for all $n \in \sN$.

Applying the above estimates to (\ref{RL5a}) and taking the supremum over $t \in [a,b]$, one obtains that
\begin{align*}
& \sup\limits_{t \in [a,b]} \left| (\T_2 \circ \gamma_n)^{-1} (\T_1(\gamma_n(t))) - (\T_2 \circ \gamma)^{-1} (\T_1(\gamma(t))) \right|
\\
& \leq L_{A,B} \sup\limits_{t \in [a,b]} \left| \T_1(\gamma_n(t)) - \T_1(\gamma(t)) \right| + L_{A',B'} \sup\limits_{\tau \in [a',b']} \left| \T_2(\gamma(\tau)) - \T_2(\gamma_n(\tau)) \right|,
\end{align*}
which tends to zero on the strength of Lemma \ref{RhomeoLem2iPol}.
\end{Proof}

We are finally ready to strengthen Proposition \ref{CITallpaths5}.
\begin{Thm}
\label{RhomeoThm}
Let $\T_1, \T_2 : \M \rightarrow \sR$ be Cauchy temporal functions. Then the map $\widetilde{\ }: C^{\sR}_{\T_1} \rightarrow C^{\sR}_{\T_2}$, defined via (\ref{Rhomeo}) is a homeomorphism.
\end{Thm}
\begin{Proof}\textbf{.}
On the strength of Proposition \ref{CITallpaths5}, we only need to show that $\gamma_n \rightarrow \gamma$ in $C^{\sR}_{\T_1}$ iff $\widetilde{\gamma}_n \rightarrow \widetilde{\gamma}$ in $C^{\sR}_{\T_2}$. In fact, showing just one of the implications is enough, because proving the other one amounts to swapping $\T_1$ with $\T_2$.

Thus, assume that $\gamma_n \rightarrow \gamma$ in $C^{\sR}_{\T_1}$ and for any $n \in \sN$ denote $\lambda_n := (\T_2 \circ \gamma_n)^{-1} \circ \T_1 \circ \gamma_n$ as well as $\lambda := (\T_2 \circ \gamma)^{-1} \circ \T_1 \circ \gamma$. Our goal is to show that $\widetilde{\gamma}_n = \gamma_n \circ \lambda_n \rightarrow \gamma \circ \lambda = \widetilde{\gamma}$ uniformly on any $[a,b] \subseteq \sR$, as this would already imply convergence in $C_{\T_2}^{\sR}$.

The sequence $(\lambda_n) \subseteq C(\sR,\sR)$ consists of strictly increasing maps, which by Lemma~\ref{RhomeoLem5} converges to the strictly increasing map $\lambda \in C(\sR,\sR)$ uniformly on compact sets. The latter implies the pointwise convergence, therefore the sequences $(\lambda_n(a))$, $(\lambda_n(b)) \subseteq \M$ are convergent and hence bounded. Denote $A := \inf_n \lambda_n(a)$ and $B := \sup_n \lambda_n(b)$. Notice that, because $\lambda$ and $\lambda_n$ for every $n \in \sN$ are continuous and strictly increasing maps, the interval $[A,B]$ contains $\lambda([a,b]) = [\lambda(a), \lambda(b)]$ as well as $\lambda_n([a,b]) = [\lambda_n(a), \lambda_n(b)]$ for all $n \in \sN$.

For any $t \in [a,b]$ one obtains
\begin{align*}
d_w\left( \gamma_n(\lambda_n(t)), \gamma(\lambda(t)) \right) & \leq d_w\left( \gamma_n(\lambda_n(t)), \gamma_n(\lambda(t)) \right) + d_w\left( \gamma_n(\lambda(t)), \gamma(\lambda(t)) \right)
\\
& \leq L_{A,B} |\lambda_n(t) - \lambda(t)| + d_w\left( \gamma_n(\lambda(t)), \gamma(\lambda(t)) \right),
\end{align*}
where $L_{A,B} > 0$ exists on the strength of Lemma \ref{RhomeoLem3}. Taking the supremum over $t \in [a,b]$, one obtains
\begin{align*}
& \sup\limits_{t \in [a,b]} d_w\left( \gamma_n(\lambda_n(t)), \gamma(\lambda(t)) \right) \leq L_{A,B} \sup\limits_{t \in [a,b]} |\lambda_n(t) - \lambda(t)| + \sup\limits_{\tau \in [\lambda(a),\lambda(b)]} d_w\left( \gamma_n(\tau), \gamma(\tau) \right).
\end{align*}
Notice now that both terms on the right-hand side tend to zero. Indeed, the rightmost term does so by assumption, whereas the other one by Lemma \ref{RhomeoLem5}.
\end{Proof}

We finish this section by studying in detail the subspaces $\I_\T \subseteq C^\sR_\T$, which were crucial in the physical discussion carried out in Section \ref{sec::discussion}.
\begin{Def}
\label{ITDef}
Let $\M$ be a globally hyperbolic spacetime and let $\T: \M \rightarrow \sR$ be a Cauchy time function. Define the space $\I_\T$ via
\begin{align*}
\I_\T := \{ \gamma \in C^\sR_\T \ | \ \T \circ \gamma = \id_\sR \}
\end{align*}
with the topology induced from $C^\sR_\T$.
\end{Def}
\begin{Rem}
\label{ITRem}
Recall that $\ev_t: C^\sR_\T \rightarrow \M$ denotes the evaluation map for any $t \in \sR$. Observe that $\I_\T = (\T \circ \ev_0)^{-1}(0) \cap (\T \circ \ev_1)^{-1}(1)$. Indeed, the inclusion $\subseteq$ is trivial, whereas to prove $\supseteq$, assume that $\gamma \in C^\sR_\T$ satisfies $\T(\gamma(0)) = 0$ and $\T(\gamma(1)) = 1$. Using (\ref{alt}), we obtain that
\begin{align*}
\forall \, t \in \sR \quad \T(\gamma(t)) = (1-t)\T(\gamma(0)) + t\T(\gamma(1)) = t,
\end{align*}
and hence $\gamma \in \I_\T$. As a side note, observe that $c_\gamma = 1$.
\end{Rem}

The above remark makes it clear that $\I_\T$ is closed in $C^\sR_\T$ and hence it is a Polish space. As it was announced in Section \ref{sec::intro}, the importance of these spaces relies on their bijectivity with the set $\cCi$ of all inextendible (hence the letter ``$\I$'') causal paths on $\M$.
\begin{Prop}
\label{IT1}
Let $\M$ and $\T$ be as above. Then the map $[ \,.\, ]: \I_\T \rightarrow \cCi$, $\gamma \mapsto [\gamma]$ is a bijection.
\end{Prop}
\begin{Proof}\textbf{.}
That the map $[ \, . \, ]$ is well defined follows from the proof of Proposition \ref{CITallpaths4}. Surjectivity is a consequence of Proposition \ref{CITallpaths3} (the last bullet), where in the formula for $\lambda$ we choose $A = 1$ and $B = 0$.

In order to prove injectivity, assume that $[\gamma_1] = [\gamma_2]$ for some $\gamma_1,\gamma_2 \in \I_\T$. This means that there exists $\lambda \in C(\sR,\sR)$ such that $\gamma_1 \circ \lambda = \gamma_2$. However, composing both sides of the last identity with $\T$, by the very definition of $\I_\T$ one obtains that
\begin{align*}
\lambda = \id_\sR \circ \lambda = \T \circ \gamma_1 \circ \lambda = \T \circ \gamma_2 = \id_\sR,
\end{align*}
and so $\gamma_1 = \gamma_2$.
\end{Proof}

Using the above bijection, one can readily topologize $\cCi$ by transporting the Polish space topology from $\I_\T$ using the bijective map $[ \, . \, ]$. Thus obtained topology turns out to be independent from the choice of $\T$, as long as we are concerning Cauchy \emph{temporal} functions.
\begin{Prop}
\label{IT2}
Let $\M$ be a globally hyperbolic spacetime and let $\T_1,\T_2: \M \rightarrow \sR$ be Cauchy temporal functions. Then the map $\widetilde{\ }: \I_{\T_1} \rightarrow \I_{\T_2}$, defined via (\ref{Rhomeo}), that is
\begin{align}
\label{IT2a}
\forall \, \gamma \in \I_{\T_1} \quad \widetilde{\gamma} := \gamma \circ \left(\T_2 \circ \gamma \right)^{-1} \circ \T_1 \circ \gamma,
\end{align}
\noindent
is a homeomorphism.
\end{Prop}
\begin{Proof}\textbf{.}
Proposition \ref{CITallpaths5} guarantees that $\widetilde{\gamma} \in C^\sR_{\T_2}$ for any $\gamma \in \I_{\T_1}$. However, noticing that $\T_2 \circ \widetilde{\gamma} = \T_1 \circ \gamma = \id_\sR$ we have that in fact $\widetilde{\gamma} \in \I_{\T_2}$ and so the map $\widetilde{\ }: \I_{\T_1} \rightarrow \I_{\T_2}$ is well defined. Swapping $\T_1$ and $\T_2$, one similarly proves that the inverse of $\widetilde{\ }$ is well defined, too.

That $\widetilde{\ }$ and its inverse are continuous follows from Theorem \ref{RhomeoThm} and from elementary properties of the subspace topology.
\end{Proof}

\section{Application in Lorentzian optimal transport \mbox{theory}}
\label{sec::proof}

The following section is devoted to proving Theorems \ref{main} \& \ref{mainvar} stated in Section \ref{sec::intro}. We begin by establishing the validity of the `easier' implication.

\begin{Proof}\textbf{ of Theorem \ref{main}, $ii) \Rightarrow i)$.} Let $\iota: J^+ \hookrightarrow \M^2$ denote the canonical topological embedding and take any $s,t \in I$ such that $s \leq t$. The map $(\ev_s, \ev_t): C_\T^I \rightarrow J^+$, $\gamma \mapsto (\gamma(s),\gamma(t))$ is well-defined and continuous, hence Borel. One can thus define $\omega_{s,t} := \left[\iota \circ (\ev_s, \ev_t)\right]_\# \sigma$, which we now show to be a causal coupling of $\mu_s$ and $\mu_t$.

Indeed, $\pi^1_\# \omega_{s,t} = [\pi^1 \circ \iota \circ (\ev_s, \ev_t)]_\# \sigma = (\ev_s)_\# \sigma = \mu_s$ and similarly $\pi^2_\# \omega_{s,t} = \mu_t$. Additionally,
\begin{align*}
\omega_{s,t}(J^+) = \sigma\left( (\ev_s, \ev_t)^{-1}(J^+) \right) = \sigma(C_\T^I) = 1,
\end{align*}
which completes the proof of $i)$.
\end{Proof}

The proof of the converse implication requires some technical preparations. Let us start by providing some auxiliary facts about the narrow topology.
\begin{Lem}
\label{properness}
Let $\X,\Y$ be Polish spaces and let $F: \X \rightarrow \Y$ be a continuous map. Then $F_{\#}: \Pf(\X) \rightarrow \Pf(\Y)$ is continuous (in the narrow topology). Moreover, if $F$ is also proper, then so is $F_{\#}$.
\end{Lem}
\begin{Proof}\textbf{.}
Suppose that $(\mu_n) \subseteq \Pf(\X)$ is narrowly convergent to $\mu \in \Pf(\X)$. Take $g \in C_b(\Y)$. One obtains
\begin{align*}
\int_{\Y} g \, d(F_{\#} \mu_n) = \int_{\X} \underbrace{(g \circ F)}_{\in \, C_b(\X)} d\mu_n \rightarrow \int_{\X} (g \circ F) d\mu = \int_{\Y} g \, d(F_\# \mu),
\end{align*}
what proves that $F_{\#}$ is continuous.

Assume now that $F$ is continuous and proper. Let $\K \subseteq \Pf(\Y)$ be compact. We want to show that $F_{\#}^{-1}(\K)$ is a compact subset of $\Pf(\X)$. By the continuity of $F_{\#}$, $F_{\#}^{-1}(\K)$ is closed, and so it suffices to additionally show that it is relatively compact. By the Prokhorov theorem (cf. \cite[Chapter 18, Theorem 17]{ModProb} or \cite[Chapter II, Theorem 6.7]{Part67}), this is, in turn, equivalent to showing that $F_{\#}^{-1}(\K)$ is \emph{tight}, i.e.
\begin{align*}
\forall \, \varepsilon > 0 \quad \exists \, \textnormal{compact } K_{\varepsilon} \subseteq \X \quad \forall \, \mu \in F_{\#}^{-1}(\K) \ \quad \mu(K_{\varepsilon}) \geq 1 - \varepsilon.
\end{align*}
Let us thus fix $\varepsilon > 0$. By assumption, $\K \subseteq \Pf(\Y)$ is compact and hence tight, therefore there exists $K^\prime_{\varepsilon} \subseteq \Y$ compact and such that $\nu( K^\prime_{\varepsilon} ) \geq 1 - \varepsilon$ for every $\nu \in \K$. In particular, this is true for $\nu = F_{\#} \mu$ for every $\mu \in F_{\#}^{-1}(\K)$. We thus obtain that
\begin{align*}
\forall \, \mu \in F_{\#}^{-1}(\K) \quad \mu\left(F^{-1}(K^\prime_{\varepsilon})\right) \geq 1 - \varepsilon.
\end{align*}
By the assumption that $F$ is proper, taking $K_{\varepsilon} := F^{-1}(K^\prime_{\varepsilon})$ completes the proof.
\end{Proof}

\begin{Lem}
\label{narrowtight}
Let $\X$ be a second-countable LCH space and suppose that the family of measures $\{\mu_n\} \subseteq \Pf(\X)$ is tight. Then $(\mu_n)$ converges to $\mu \in \Pf(\X)$ narrowly iff $\ \forall \, f \in C_c(\X) \ \int_\X f d\mu_n \rightarrow \int_\X f d\mu$.
\end{Lem}
\begin{Proof}\textbf{.}
The ``$\Rightarrow$'' part is trivial. To prove the ``$\Leftarrow$'' part, recall that every second countable LCH space $\X$ admits an \emph{exhaustion by compact sets}, i.e. a sequence $(\K_m)$ of compact subsets of $\X$ such that $\K_m \subseteq \textnormal{int} \, \K_{m+1}$ for all $m$ and $\bigcup_{m \in \sN} \K_m = \X$. Furthermore, let $(\varphi_m) \subseteq C_c(\X)$ be a sequence of functions satisfying $0 \leq \varphi_m \leq 1$, $\varphi_m|_{\K_m} \equiv 1$ and $\supp \, \varphi_m \subseteq \K_{m+1}$, existing by Urysohn's lemma. Take any $g \in C_b(\X)$. Then, by assumption,
\begin{align*}
\forall \, m \in \sN \quad \lim\limits_{n \rightarrow +\infty} \int_\X g \varphi_m d\mu_n = \int_\X g \varphi_m d\mu.
\end{align*}
Since $\X$ is Polish, $\mu$ is inner regular, which, together with the tightness of $\{\mu_n\}$, means that
\begin{align*}
\forall \, \varepsilon > 0 \ \exists \textnormal{ compact } K \subseteq \X \quad \mu(K^c) \leq \varepsilon \quad \textnormal{and} \quad \forall \, n \in \sN \quad \mu_n(K^c) \leq \varepsilon.
\end{align*}
We claim that $\int_\X g d\mu_n \rightarrow \int_\X g d\mu$. Indeed, fix $\varepsilon > 0$ and take $K \subseteq \X$ compact and such that
\begin{align*}
\mu(K^c) \leq \frac{\varepsilon}{4\|g\|} \quad \textnormal{and} \quad \forall \, n \in \sN \quad \mu_n(K^c) \leq \frac{\varepsilon}{4\|g\|},
\end{align*}
where $\|g\|$ denotes the supremum of $g$. We now have that
\begin{align*}
\left| \int_\X g d\mu_n - \int_\X g d\mu \right| & = \left| \int_\X g \varphi_m d\mu_n - \int_\X g \varphi_m d\mu + \int_{\K_m^c} g (1 - \varphi_m) d\mu_n - \int_{\K_m^c} g (1 - \varphi_m) d\mu \right|
\\
& \leq \left| \int_\X g \varphi_m d\mu_n - \int_\X g \varphi_m d\mu \right| + \|g\| \mu_n(\K_m^c) + \|g\| \mu(\K_m^c).
\end{align*}
It now remains to take $m$ large enough to have the inclusion $K \subseteq \K_m$ (which makes the two rightmost summands less than $\varepsilon/4$ each) and then choose $N \in \sN$ such that the leftmost summand falls below $\varepsilon/2$ for all $n > N$.
\end{Proof}

In the following, let $\itPi(\mu,\nu)$ ($\itPi_c(\mu,\nu)$) denote the set of all (causal) couplings of the measures $\mu$ and  $\nu$ (cf. Definition \ref{JGPcausality_def_true}).
\begin{Lem}
\label{Pi_cCompact}
Let $\M$ be a causally simple spacetime and let $\mu,\nu \in \Pf(\M)$. Then $\itPi_c(\mu,\nu)$ is a narrowly compact subset of $\Pf(\M^2)$.
\end{Lem}
\begin{Proof}\textbf{.}
It is well known that $\itPi(\mu,\nu)$ is narrowly compact in $\Pf(\M^2)$ \cite[Theorem~1.5]{UsersGuide}. Hence we only need to show that $\itPi_c(\mu,\nu)$ is closed in $\itPi(\mu,\nu)$. To this end, take any sequence $(\omega_n) \subseteq \itPi_c(\mu,\nu)$ convergent to some $\omega \in \itPi(\mu,\nu)$. We need to prove that \mbox{$\omega(J^+) = 1$}. This, however, is a direct consequence of the portmanteau theorem (cf. \cite[Chapter 18, Theorem 6]{ModProb} or \cite[Chapter II, Theorem 6.1]{Part67}), on the strength of which
\begin{align*}
\omega_n \rightarrow \omega \textnormal{ narrowly } \ \Leftrightarrow \quad \forall \, \textnormal{ closed } C \subseteq \M^2 \quad \limsup\limits_{n \rightarrow +\infty} \omega_n(C) \leq \omega(C).
\end{align*}
Taking $C = J^+$ --- which is closed by the causal simplicity of $\M$ --- we easily obtain
\begin{align*}
1 = \limsup\limits_{n \rightarrow +\infty} \omega_n(J^+) \leq \omega(J^+) \leq 1,
\end{align*}
and so $\omega(J^+) = 1$.
\end{Proof}

The next Proposition is an analogue of \cite[Proposition 3.3]{Suhr2016}.
\begin{Prop}
\label{Borelinverse}
Let $\M$ be a globally hyperbolic spacetime and let $\T: \M \rightarrow \sR$ be a time function. For any $a,b \in \sR$ the map $(\ev_a, \ev_b): C^{[a,b]}_\T \rightarrow J^+$ is proper and admits a Borel right inverse.
\end{Prop}
\begin{Proof}\textbf{.}
To prove properness, let $\K \subseteq J^+$ be compact and observe that
\begin{align*}
(\ev_a, \ev_b)^{-1}(\K) \subseteq (\ev_a, \ev_b)^{-1}\left(\pi^1(\K) \times \pi^2(\K) \, \cap \, J^+\right) = C^{[a,b]}_\T(\pi^1(\K),\pi^2(\K)),
\end{align*}
which, on the strength of the continuity of the map $(\ev_a, \ev_b)$ and Proposition \ref{Leray1}, means that $(\ev_a, \ev_b)^{-1}(\K)$ is a closed subset of a compact set and hence it is itself compact.

For the second part of the proposition's statement we use the standard measurable selection result, by which a continuous map from a $\sigma$-compact metrizable space onto a metrizable space admits a Borel right inverse \cite[Corollary I.8]{Fabec}.

Both $C^{[a,b]}_\T$ and $J^+$ are metrizable spaces, and the latter can be shown to be $\sigma$-compact \cite[Section 3]{EcksteinMiller2015}. Since the map $(\ev_a, \ev_b)$ is proper, we have that $C^{[a,b]}_\T = (\ev_a, \ev_b)^{-1}(J^+)$ is $\sigma$-compact as well. In order to prove that the continuous map $(\ev_a, \ev_b)$ is surjective, observe that, by the very definition of the causal precedence relation, for any pair of events $(p,q) \in J^+$ there exists a causal curve connecting them, which by Proposition \ref{CITallpaths} can be reparametrized so that it becomes an element of $C^{[a,b]}_\T$.
\end{Proof}

Given a pair of continuous (causal) curves $\gamma_1: [a,b] \rightarrow \M$, $\gamma_2: [b,c] \rightarrow \M$ such that $\gamma_1(b) = \gamma_2(b)$, one can easily concatenate them, obtaining another continuous (causal) curve $\gamma_1 \sqcup \gamma_2: [a,c] \rightarrow \M$ through the obvious piecewise definition. In the proof of Theorem \ref{main}, however, we will need a way to ``concatenate'' two \emph{measures} on spaces of curves, one on $C^{[a,b]}_\T$ and the other on $C^{[b,c]}_\T$. To this end, recall first the \emph{disintegration theorem} \cite[Theorem 5.3.1]{ambrosio2008gradient}.
\begin{Thm}
Let $\X, \Y$ be Polish spaces, $\mu \in \Pf(\Y)$ and let $\pi: \Y \rightarrow \X$ be a Borel map. Denote $\nu := \pi_\# \mu$. Then there exists a $\nu$-a.e. uniquely determined family of probability measures $\{\mu^x\}_{x\in\X} \subseteq \Pf(\Y)$ such that
\begin{itemize}
\item For any Borel subset $E \subseteq \Y$ the map $\X \ni x \mapsto \mu^x(E)$ is Borel.
\item Measures $\mu^x$ live on the fibers of $\pi$, that is $\mu^x(\Y \setminus \pi^{-1}(x)) = 0$ for $x \in \X$ $\nu$-a.e.
\item For any Borel map $f: \Y \rightarrow [0,+\infty]$
\begin{align}
\label{disintegration}
\int_\Y f d\mu = \int_\X \left( \int_{\pi^{-1}(x)} f(y) d\mu^x(y) \right) d\nu(x).
\end{align}
\end{itemize}
We call the family $\{\mu^x\}_{x\in\X}$ the \emph{disintegration} of $\mu$ with respect to (w.r.t.) $\pi$.
\end{Thm}

\begin{Def}
\label{concatDef}
Let $\M$ be a stably causal spacetime and let $\T: \M \rightarrow \sR$ be a time function. For any fixed $a,b,c \in \sR$, $a < b < c$ let $\Y := \{ (\gamma_1, \gamma_2) \in C^{[a,b]}_\T \times C^{[b,c]}_\T \ | \ \gamma_1(b) = \gamma_2(b) \}$ denote the (Polish) space of concatenable pairs of curves. Consider the concatenation map $\sqcup: \Y \rightarrow C^{[a,c]}_\T$, which is obviously continuous and hence Borel. For any $\sigma_1 \in \Pf(C^{[a,b]}_\T)$ and $\sigma_2 \in \Pf(C^{[b,c]}_\T)$, which are \emph{compatible} in the sense that $(\ev_b)_\# \sigma_1 = (\ev_b)_\# \sigma_2 =: \nu$, we define their \emph{concatenation} $\sigma_1 \sqcup \sigma_2 \in \Pf(C^{[a,c]}_\T)$ with the help of the Riesz--Markov--Kakutani representation theorem via
\begin{align}
\label{measure_concatenation}
\int_{C^{[a,c]}_\T} F d(\sigma_1 \sqcup \sigma_2) := \int_\M \left( \int_\Y F( \gamma_1 \sqcup \gamma_2 ) d(\sigma_1^x \times \sigma_2^x )(\gamma_1, \gamma_2) \right) d\nu(x)
\end{align}
for any $F \in C_c(C^{[a,c]}_\T)$, where $\{\sigma_i^x\}_{x \in \M}$ is the disintegration of $\sigma_i$ w.r.t. $\ev_b$ for $i=1,2$.
\end{Def}

\begin{Rem}
In order to convince oneself that the integral on the right-hand side of (\ref{measure_concatenation}) is well defined, introduce a bounded Borel map $\Phi: C^{[a,b]}_\T \times C^{[b,c]}_\T \rightarrow \sR$ via $\Phi(\gamma_1,\gamma_2) := F(\gamma_1 \sqcup \gamma_2)$ for $(\gamma_1,\gamma_2) \in \Y$ and zero otherwise. Observe that the integral can now be rewritten, by Fubini's theorem, as
\begin{align*}
\int_\M \left( \int_\Y (F \circ \sqcup) d(\sigma_1^x \times \sigma_2^x ) \right) d\nu(x) = \int_\M \left( \int_{C^{[b,c]}_\T} \left( \int_{C^{[a,b]}_\T} \Phi(\gamma_1,\gamma_2) d\sigma^x_1(\gamma_1) \right) d\sigma^x_2(\gamma_2) \right) d\nu(x),
\end{align*}
where the map $(x,\gamma_2) \mapsto \int_{C^{[a,b]}_\T} \Phi(\gamma_1,\gamma_2) d\sigma^x_1(\gamma_1)$ is Borel and bounded by the definition of the disintegration, and so is the map $x \mapsto \int_{C^{[b,c]}_\T} \left( \int_{C^{[a,b]}_\T} \Phi(\gamma_1,\gamma_2) d\sigma^x_1(\gamma_1) \right) d\sigma^x_2(\gamma_2)$.
\end{Rem}

\begin{Rem}
\label{concatRem1}
In the discussion preceding the disintegration theorem, as well as in Definition \ref{concatDef} everything is still valid if we replace $[a,b]$ with $(a,b]$ or even with $(-\infty,b]$ or if we replace $[b,c]$ with $[b,c)$ or even with $[b,+\infty)$. In other words, nothing prevents from concatenating measures living on the spaces of noncompact curves.
\end{Rem}
\begin{Rem}
\label{concatRem2}
As one would expect, it is true (and easy to check) that
\begin{align*}
(\ev_t)_\#(\sigma_1 \sqcup \sigma_2) = \left\{\begin{array}{ll}
                                            (\ev_t)_\# \sigma_1 & \textrm{for } t < b
                                            \\
                                            \nu & \textrm{for } t = b
                                            \\
                                            (\ev_t)_\# \sigma_2 & \textrm{for } t > b
                                            \end{array}\right. .
\end{align*}
In fact, this is the reason why we have chosen the symbol $\sqcup$ to denote the concatenation operation in the first place.
\end{Rem}

We now prove that every causal time-evolution of measures is automatically narrowly continuous.
\begin{Prop}
\label{caus_is_cont}
Let $\M$ be a globally hyperbolic spacetime and let $\T: \M \rightarrow \sR$ be a Cauchy temporal function. Consider a map $\mu: I \rightarrow \Pf(\M)$, $t \mapsto \mu_t$ such that $\supp \, \mu_t \subseteq \T^{-1}(t)$ for every $t \in I$. If the map $\mu$ is causal, then it is narrowly continuous.
\end{Prop}
\begin{Proof}\textbf{.}
Fix any $a,b \in I$ with $a < b$. First, let us show that the family $\{\mu_t\}_{t \in [a,b]}$ is tight. By Lemma \ref{narrowtight}, this will allow us to use only the compactly supported test functions when proving the narrow continuity of the restricted map $\mu|_{[a,b]}$.

Indeed, fix $\varepsilon > 0$ and take $K_a \subseteq \T^{-1}(a)$ compact and such that $\mu_a(K_a) \geq 1 - \varepsilon$. Define $K := J^+(K_a) \cap J^-(\T^{-1}(b))$, which is compact on the strength of Proposition \ref{PropCCC} (see \hyperref[sec::preliminaries]{Appendix}). Of course, $\supp \, \mu_t \subseteq \T^{-1}(t) \subseteq J^-(\T^{-1}(b))$ for all $t \in [a,b]$ and hence
\begin{align*}
\mu_t(K) = \mu_t(J^+(K_a) \cap J^-(\T^{-1}(b))) = \mu_t(J^+(K_a)) \geq \mu_a(J^+(K_a)) = \mu_a(K_a) \geq 1 - \varepsilon,
\end{align*}
where the inequality follows from (\ref{causal_measure_map}) and the following characterization of $\preceq$ proven in \cite{EcksteinMiller2015} valid for causally simple spacetimes
\begin{align*}
\mu \preceq \nu \quad \Leftrightarrow \quad \forall \ \textrm{compact } C \subseteq \M \quad \mu(J^+(C)) \leq \nu(J^+(C)).
\end{align*}
This completes the proof of tightness.

We now move on to showing that $\lim_{s \rightarrow 0^+} \mu_{t+s} = \mu_t$ for any fixed $t \in [a,b)$ (for the other one-sided limit the proof is analogous).

Begin by fixing $\omega_{t,s} \in \itPi_c(\mu_t,\mu_{t+s})$ for each $s \in (0, b-t]$. For any $f \in C_c(\M)$ one has
\begin{align}
\label{caus_is_cont1}
\left| \int_\M f \, d\mu_t - \int_\M f \, d\mu_{t+s} \right| & = \left| \int_{\M^2} (f(p) - f(q)) d\omega_{t,s}(p,q) \right|
\\
\nonumber
& \leq \int_{\supp \, \omega_{t,s}} |f(p) - f(q)| d\omega_{t,s}(p,q).
\end{align}

Our aim is to prove that the rightmost integral becomes arbitrarily small for $s$ sufficiently close to zero. To this end, let us first prove the following bound similar to the one appearing in the proof of Lemma \ref{RhomeoLem3}:
\begin{align}
\label{caus_is_cont2}
\forall \, (p,q) \in \supp \, \omega_{t,s} \quad d_w(p,q) \leq s \max\limits_{r \in J^+(p) \cap J^-(q)} \sqrt{2u(r)\alpha(r)},
\end{align}
where $d_w$ again denotes the distance function associated with the complete Riemannian metric $w$ given by (\ref{Rmetric2}).

Indeed, observe that
\begin{align}
\label{caus_is_cont3}
\supp \, \omega_{t,s} \subseteq \left[ \T^{-1}(t) \times \T^{-1}(t+s) \right] \cap J^+,
\end{align}
therefore $\T(p) = t$, $\T(q) = t+s$ and $p \preceq q$. Let thus $\gamma: [t,t+s] \rightarrow \M$ be a future-directed causal curve connecting $p$ with $q$ parametrized so as to make it an element of $C^{[t,t+s]}_\T$ (the existence of such $\gamma$ is guaranteed by the definition of $\preceq$ and by Proposition \ref{CITallpaths}). Reasoning similarly as in the proof of Lemma \ref{RhomeoLem3}, we obtain
\begin{align*}
d_w(p,q) & = d_w(\gamma(t),\gamma(t+s)) \leq c_\gamma \int_t^{t+s} \sqrt{2u(\gamma(\tau)) \cdot \alpha(\gamma(\tau))} d\tau
\\
& \leq c_\gamma \max\limits_{r \in J^+(p) \cap J^-(q)} \sqrt{2u(r)\alpha(r)} \cdot s,
\end{align*}
where noticing that $c_\gamma = (\T(q) - \T(p))/s = 1$ finishes the proof of (\ref{caus_is_cont2}). This bound, however, has the downside of being dependent on $s$ and $t$ (through $p$ and $q$). It would be desirable to replace there the set $J^+(p) \cap J^-(q)$, over which the maximum is evaluated, with a compact set independent from $p,q$. One way of achieving this is to notice that when estimating the rightmost integral in (\ref{caus_is_cont1}), we can restrict to $(p,q) \not\in (K_f^c)^2$, where $K_f := \supp \, f$ is compact. We are thus looking for a \emph{compact} superset of $\supp \, \omega_{t,s} \setminus (K_f^c)^2$, which would be independent from $s,t$.

One possibility is
\begin{align}
\label{KDef}
\K := K_f \times \left[ J^+(K_f) \cap J^-(\T^{-1}(b)) \right] \ \cup \ \left[ J^-(K_f) \cap J^+(\T^{-1}(a)) \right] \times K_f,
\end{align}
whose compactness is guaranteed by the global hyperbolicity of $\M$ (\hyperref[sec::preliminaries]{Appendix}, Proposition \ref{PropCCC} $iii)$). Furthermore, using (\ref{caus_is_cont3}), one can easily check that $\K$ is indeed a superset of $\supp \, \omega_{t,s} \setminus (K_f^c)^2$.

Ultimately, since $(p,q) \in \K$ implies that $J^+(p) \cap J^-(q) \subseteq J^+(\pi^1(\K)) \cap J^-(\pi^2(\K))$ (notice the latter set is still compact by Proposition \ref{PropCCC} $i)$), we obtain from (\ref{caus_is_cont2}) a somewhat modified bound
\begin{align}
\label{caus_is_cont4}
\forall \, (p,q) \in \supp \, \omega_{t,s} \setminus (K_f^c)^2 \quad \ d_w(p,q) \leq s \max\limits_{r \in J^+(\pi^1(\K)) \cap J^-(\pi^2(\K))} \sqrt{2u(r)\alpha(r)},
\end{align}
where the maximum is now clearly independent from both $s$ and $t$.

Coming back to (\ref{caus_is_cont1}), we will show that
\begin{align}
\label{caus_is_cont5}
\lim\limits_{s \rightarrow 0^+} \int_{\supp \, \omega_{t,s} \setminus (K_f^c)^2} |f(p) - f(q)| d\omega_{t,s}(p,q) = 0.
\end{align}
To this end, fix $\varepsilon > 0$ and observe that (by the Heine--Borel theorem) $f \in C_c(\M)$ is uniformly continuous, what means that there exists $\delta > 0$ such that $d_w(p,q) < \delta \ \Rightarrow \ |f(p) - f(q)| < \varepsilon$ for any $p,q \in \M$.

Let us thus consider $s < \delta \cdot \left(\max_{r \in J^+(\pi^1(\K)) \cap J^-(\pi^2(\K))} \sqrt{2u(r)\alpha(r)}\right)^{-1}$. We obtain that
\begin{align*}
\int_{\supp \, \omega_{t,s} \setminus (K_f^c)^2} |f(p) - f(q)| d\omega_{t,s}(p,q) < \varepsilon \int_{\supp \, \omega_{t,s} \setminus (K_f^c)^2} d\omega_{t,s}(p,q) \leq \varepsilon,
\end{align*}
what completes the proof that $\lim_{s \rightarrow 0^+} \mu_{t+s} = \mu_t$ for any $t \in [a,b)$. One similarly shows that $\lim_{s \rightarrow 0^-} \mu_{t+s} = \mu_t$ for any $t \in (a,b]$, and hence we obtain that the map $\mu$ is continuous on $[a,b]$. But the latter was an arbitrary compact subinterval of $I$, therefore $\mu$ is in fact continuous on the entire $I$.
\end{Proof}

We are finally ready to prove the implication $i) \Rightarrow ii)$ of Theorem \ref{main}. We shall do it in four steps: first for $I = [a,b]$ (for some $a,b \in \sR, a<b$), then for $I = [0,+\infty)$, and afterwards for $I = \sR$, which altogether will imply the theorem's statement for any interval.

\begin{Proof} \textbf{of Theorem \ref{main}, $i) \Rightarrow ii)$.}

\textbf{Step 1. The $I = [a,b]$ case.} The idea is to construct a sequence $(\sigma_n) \subseteq \Pf(C^{[a,b]}_\T)$ such that $(\ev_t)_\# \sigma_n = \mu_t$ for all $t$ of the form $t^{n}_i := a + (b-a)i/2^n$, $i=0,1,2,3,\ldots,2^n$ and then show that it has a convergent subsequence, whose limit $\sigma$ satisfies the above identity for \emph{any} $t \in [a,b]$.

To this end, fix $n \in \sN$ and for any $i = 1, \ldots, 2^n$ let $S^i: J^+ \rightarrow C_\T^{[t^n_{i-1},t^n_i]}$ denote the Borel map such that $\left( \ev_{t^{n}_{i-1}}, \ev_{t^{n}_i} \right) \circ S^i = \textrm{id}_{J^+}$, existing by Proposition \ref{Borelinverse}. Furthermore, for any $i = 1, \ldots, 2^n$ let $\omega_i$ be a causal coupling of $\mu_{t^{n}_{i-1}}$ and $\mu_{t^{n}_i}$, existing by $i)$. Observe that because $\omega_i(J^+) = 1$, therefore $\omega_i|_{\Bf(J^+)} \in \Pf(J^+)$ for all $i = 1, \ldots, 2^n$, where $\Bf(J^+)$ denotes the $\sigma$-algebra of Borel subsets of $J^+$. All this allows us to define $\sigma_n$ via multiple concatenation (Definition \ref{concatDef}) as
\begin{align}
\label{mainproof1}
\sigma_n := S^1_\# \left(\omega_1|_{\Bf(J^+)}\right) \sqcup S^2_\# \left(\omega_2|_{\Bf(J^+)}\right) \sqcup S^3_\# \left(\omega_3|_{\Bf(J^+)}\right) \sqcup \ldots \sqcup S^{2^n}_\# \left(\omega_{2^n}|_{\Bf(J^+)}\right).
\end{align}
Observe that the right-hand side of (\ref{mainproof1}) is unambiguous, because the operation $\sqcup$ can be easily proven to be associative.

Notice now that $[\iota \circ (\ev_a, \ev_b)]_\# \sigma_n$ is a causal coupling of $\mu_a$ and $\mu_b$, what can be shown completely analogously as in the proof of the implication $ii) \Rightarrow i)$ above. By the arbitrariness of $n$, we thus obtain that $(\sigma_n) \subseteq [\iota \circ (\ev_a, \ev_b)]_\#^{-1}(\itPi_c(\mu_a,\mu_b))$.

We now make the crucial observation: the set $[\iota \circ (\ev_a, \ev_b)]_\#^{-1}(\itPi_c(\mu_a,\mu_b))$ is compact on the strength of Proposition \ref{Borelinverse}, Lemma \ref{Pi_cCompact} and the obvious fact that $\iota: J^+ \hookrightarrow \M^2$ is a proper map. Therefore, $(\sigma_n)$ has a subsequence that narrowly converges to certain $\sigma \in \Pf(C^{[a,b]}_\T)$.

Observe that, by the very construction of the sequence $(\sigma_n)$, it is true that $(\ev_{t^{n}_i})_\# \sigma = \mu_{t^{n}_i}$ for any $n \in \sN$ and any $i=0,1,\ldots,2^n$. However, by Proposition \ref{caus_is_cont}, this already implies that $(\ev_t)_\# \sigma = \mu_t$ is actually true for all $t \in [a,b]$.
\\

\textbf{Step 2. The $I = [0,+\infty)$ case.} For the sake of convenience, for any $i \in \sN$ denote $\X_i := C^{[i-1,i]}_\T(\T^{-1}(i-1), \T^{-1}(i))$ and let $\sigma_i \in \Pf(\X_i)$ satisfy $(\ev_t)_\#\sigma_i = \mu_t$ for every $t \in [i-1,i]$. Now, for any $n \in \sN$ define a measure $\bsigma_n \in \Pf\left(\prod_{i=1}^n \X_i\right)$ recursively as
\begin{align}
\label{mainproof2}
\bsigma_1 := \sigma_1 \qquad \textrm{and} \qquad \forall \, n \in \sN \ \ \bsigma_{n+1} := \int_{\T^{-1}(n)} \left( \bsigma_n^x \times \sigma_{n+1}^x \right) d\mu_n(x),
\end{align}
where:
\begin{itemize}
\item $\{\bsigma_n^x\}_{x \in \T^{-1}(n)}$ is the disintegration of $\bsigma_n$ w.r.t. the map $\ev_n \circ \pi^n: \prod_{i=1}^n \X_i \rightarrow \T^{-1}(n)$,
\item $\{\sigma_{n+1}^x\}_{x \in \T^{-1}(n)}$ is the disintegration of $\sigma_{n+1}$ w.r.t. the map $\ev_n: \X_{n+1} \rightarrow \T^{-1}(n)$.
\end{itemize}
For later use, we need to prove the following three properties of $\bsigma_n$'s
\begin{align}
\label{mainproof3}
& \forall \, n \in \sN \quad \bpi^n_\# \bsigma_{n+1} = \bsigma_n \qquad \textrm{and} \qquad \pi^n_\# \bsigma_n = \sigma_n,
\\
\label{mainproof4}
& \forall \, n \in \sN \quad \supp \, \bsigma_n \subseteq \left\{ (\gamma_i) \in \prod_{i = 1}^n \X_i \ | \ \gamma_i(i) = \gamma_{i+1}(i), \ i=1,\ldots,n-1 \right\},
\end{align}
where $\bpi^{n}$ denotes the canonical projection on the \emph{first $n$ arguments} (not to be confused with $\pi^{n}$, which projects on the \emph{$n$-th coordinate}).

Identities (\ref{mainproof3},\ref{mainproof4}) can be proven by a direct computation. Indeed, for any Borel map $F: \prod_{i = 1}^n \X_i \rightarrow [0,+\infty]$ we have
\begin{align*}
& \int_{\prod_{i = 1}^n \X_i} F \, d(\bpi^n_\# \bsigma_{n+1})
\\
& = \int_{\T^{-1}(n)} \left( \int_{\prod_{i = 1}^{n+1} \X_i} F(\gamma_1, \ldots, \gamma_n) d(\bsigma_n^x \times \sigma_{n+1}^x)(\gamma_1,\ldots,\gamma_n,\gamma_{n+1}) \right) d\mu_n(x)
\\
&  = \int_{\T^{-1}(n)} \left( \int_{\prod_{i = 1}^{n} \X_i} F\, d\bsigma_n^x \right) d\mu_n(x) = \int_{\prod_{i = 1}^n \X_i} F \, d\bsigma_n.
\end{align*}
Similarly, for any Borel map $f: \X_n \rightarrow [0,+\infty]$, assuming $n \geq 2$ (the case $n=1$ is trivial),
\begin{align*}
& \int_{\X_n} f \, d(\pi^n_\# \bsigma_n) = \int_{\T^{-1}(n-1)} \left( \int_{\prod_{i = 1}^{n} \X_i} f(\gamma_n) d(\bsigma_{n-1}^x \times \sigma_n^x)(\gamma_1,\ldots,\gamma_{n-1},\gamma_n) \right) d\mu_n(x)
\\
& = \int_{\T^{-1}(n-1)} \left( \int_{\X_n} f \, d\sigma_n^x \right) d\mu_n(x) = \int_{\X_n} f \, d\sigma_n.
\end{align*}

We now proceed to proving (\ref{mainproof4}) by induction over $n$.

For $n=1$ there is nothing to prove. Assume then that (\ref{mainproof3}) is proven up to a certain $n$ and suppose that $(\gamma_i) \in \supp \, \bsigma_{n+1}$, but nevertheless there exists $i_0 \in \{1,\ldots,n\}$ such that $\gamma_{i_0}(i_0) \neq \gamma_{i_0+1}(i_0)$.

If $i_0 < n$, then $\bpi^n((\gamma_i)) \not\in \supp \, \bsigma_n$ by the induction hypothesis. On the other hand, using (\ref{mainproof3}) one can easily show that $\bpi^n(\supp \, \bsigma_{n+1}) \subseteq \supp \, \bsigma_n$ and we obtain a contradiction.

Thus, the only remaining possibility is $i_0 = n$. Denote $p := \gamma_n(n)$ and $q := \gamma_{n+1}(n)$. By assumption $p \neq q$ and so there exist their open neighbourhoods $U_p, U_q \subseteq \M$, which are disjoint. Define an open neighbourhood of $(\gamma_i)$ in $\prod_{i = 1}^{n+1} \X_i$ via
\begin{align*}
\U := (\ev_n \circ \pi^n)^{-1}(U_p) \times \ev_n^{-1}(U_q).
\end{align*}
By assumption, $\bsigma_{n+1}(\U) > 0$. On the other hand, one has that
\begin{align}
\label{mainproof5}
(\bsigma_n^x \times \sigma_{n+1}^x)(\U) = 0 \quad \textrm{ for any } x \in \T^{-1}(n) \textrm{ for which it is defined.}
\end{align}
Indeed, by the disintegration theorem, for any $x$ as specified above,
\begin{align*}
\supp \, \left(\bsigma_n^x \times \sigma_{n+1}^x\right) \subseteq \supp \, \bsigma_n^x \times \supp \, \sigma_{n+1}^x \subseteq (\ev_n \circ \pi^n)^{-1}(x) \times \ev_n^{-1}(x),
\end{align*}
and observe that the rightmost set is disjoint with $\U$, as otherwise we would have $x \in U_p \cap U_q$. By (\ref{mainproof2}), this means that $\bsigma_{n+1}(\U) = 0$, hence a contradiction.

Coming back to the main course of the proof, we now invoke the Kolmogorov extension theorem as given in \cite{ambrosio2008gradient}. Suppose one is given a family of Polish spaces $\{ \X_i \}_{i \in \sN}$ and a family of measures $\{\bsigma_n\}_{n \in \sN}$ such that $\bsigma_n \in \Pf(\prod_{i = 1}^n \X_i)$ and satisfying $\bpi^n_\# \bsigma_{n+1} = \bsigma_n$ (which, by (\ref{mainproof2},\ref{mainproof3}), is the case here). Then there exists $\bsigma_{\infty} \in \Pf\left(\prod_{i = 1}^\infty \X_i\right)$ such that $\bpi^n_\# \bsigma_{\infty} = \bsigma_n$.

Of course, $\bsigma_{\infty}$ is \emph{not} the desired measure $\sigma \in \Pf(C^{[0,+\infty)}_\T)$, however we will now construct the latter from the former. To this end, define the map $H: C^{[0,+\infty)}_\T \rightarrow \prod_{i = 1}^\infty \X_i$ via
\begin{align}
\label{homeo}
\forall \, \gamma \in C^{[0,+\infty)}_\T \quad H(\gamma) := (\gamma|_{[i-1,i]})_{i \in \sN}.
\end{align}
Observe that the image $H\left(C^{[0,+\infty)}_\T\right)$ is the set of those sequences $(\gamma_i) \in \prod_{i = 1}^\infty \X_i$ which satisfy $\gamma_i(i) = \gamma_{i+1}(i)$ for \emph{all} $i \in \sN$.

Clearly, $H$ is one-to-one and hence it has an inverse
\begin{align}
\label{Hinverse}
H^{-1}: H\left(C^{[0,+\infty)}_\T\right) \rightarrow C^{[0,+\infty)}_\T, \qquad H^{-1}((\gamma_i))(t) = \gamma_{\lfloor t \rfloor + 1}(t)
\end{align}
for any $t \geq 0$. Intuitively, $H^{-1}$ can be regarded as a simultaneous concatenation of countably many causal curves. We claim that both $H$ and $H^{-1}$ are \emph{continuous} maps.

To prove this claim, let us denote $\bgamma := (\gamma_i) \in \prod_{i = 1}^\infty \X_i$ and recall that a sequence $(\bgamma_n) \subseteq \prod_{i = 1}^\infty \X_i$ is convergent in the product topology iff for every $i \in \sN$ the sequence $(\pi^i(\bgamma_n)) \subseteq \X_i$ is convergent. Now, take any $(\bgamma_n) \subseteq H\left(C^{[0,+\infty)}_\T\right)$ convergent in the product topology, and consider a sequence $(H^{-1}(\bgamma_n)) \subseteq C^{[0,+\infty)}_\T$. Taking any compact subset of $[0,+\infty)$, we can cover it with finitely many intervals of the form $[i-1,i]$, on all of which we have the uniform convergence of sequences $(H^{-1}(\bgamma_n)|_{[i-1,i]})$ and hence $(H^{-1}(\bgamma_n))$ converges in the compact-open topology. Conversely, if $(\gamma_n) \subseteq C^{[0,+\infty)}_\T$ converges in the compact-open topology, then $(\gamma_n|_{[i-1,i]})$ converges in $\X_i$ for every $i \in \sN$ and hence $(H(\gamma_n))$ converges in the product topology.

Notice that the above reasoning proves in particular that the image of $H$ is closed in $\prod_{i = 1}^\infty \X_i$ and as such it is a Polish space.

To finish the proof, we want to define $\sigma := (H^{-1})_\# \bsigma_{\infty}$, however we do not a priori know whether $\bsigma_{\infty}$ is concentrated on the image of $H$, i.e. whether
\begin{align}
\label{mainproof6}
\supp \, \bsigma_{\infty} \subseteq H\left(C^{[0,+\infty)}_\T\right) = \left\{ (\gamma_i) \in \prod_{i = 1}^{\infty} \X_i \ | \ \gamma_i(i) = \gamma_{i+1}(i), \ i \in \sN \right\}.
\end{align}
This is, in a sense, the ``$n \rightarrow +\infty$'' version of property (\ref{mainproof4}) and the reasoning is somewhat similar. Concretely, suppose on the contrary that one can find $\bgamma = (\gamma_i) \in \supp \, \bsigma_{\infty}$, for which there exists $i_0 \in \sN$ such that $\gamma_{i_0}(i_0) \neq \gamma_{i_0+1}(i_0)$. This property will be preserved if we truncate $\bgamma$ to $\bpi^n(\bgamma)$ for any fixed $n \geq i_0 + 1$ and therefore $\bpi^n(\bgamma) \not\in \supp \, \bsigma_n$. On the other hand, using the property that $\bpi^n_\# \bsigma_{\infty} = \bsigma_n$ guaranteed by the Kolmogorov theorem, one can show that $\bpi^n(\supp \, \bsigma_{\infty}) \subseteq \supp \, \bsigma_n$ and thus we arrive at a contradiction.

All in all, we finally have a well-defined $\sigma \in \Pf(C^{[0,+\infty)}_\T)$. One can still be anxious whether it really inherits the property $(\ev_t)_\# \sigma = \mu_t$ ($t \geq 0$) from its ``constituents'' $\sigma_i$'s. This can be in fact checked directly by first noticing that, by formula (\ref{Hinverse}),
\begin{align*}
\forall \, t \geq 0 \quad \ev_t \circ H^{-1} = \ev_t \circ \pi^{\lfloor t \rfloor + 1} = \ev_t \circ \pi^{\lfloor t \rfloor + 1} \circ \bpi^{\lfloor t \rfloor + 1},
\end{align*}
and then by applying this observation as follows:
\begin{align*}
(\ev_t)_\# \sigma = (\ev_t \circ H^{-1})_\# \bsigma_{\infty} = \left( \ev_t \circ \pi^{\lfloor t \rfloor + 1} \circ \bpi^{\lfloor t \rfloor + 1} \right)_\# \bsigma_{\infty} = ( \ev_t )_\# \sigma_{\lfloor t \rfloor + 1} = \mu_t
\end{align*}
for any $t \geq 0$, where we have used both properties (\ref{mainproof3},\ref{mainproof4}).
\\

\textbf{Step 3. The $I = \sR$ case.} Let $\sigma_+ \in \Pf(C^{[0,+\infty)}_\T)$ denote the measure constructed in Step 2. One can similarly construct $\sigma_- \in \Pf(C^{(-\infty,0]}_\T)$ such that $(\ev_t)_\# \sigma_- = \mu_t$ for all $t \leq 0$. To this end, define $\X_i := C^{[-i,-i+1]}_\T(\T^{-1}(-i), \T^{-1}(-i+1))$ and proceed as before, applying some minor modifications reflecting the fact that here, while constructing $\bsigma_{\infty}$, we are moving towards the \emph{lower} values of $\T$.

On the strength of Remark \ref{concatRem1}, one can define $\sigma := \sigma_- \sqcup \sigma_+$, and Remark \ref{concatRem2} guarantees that $(\ev_t)_\# \sigma = \mu_t$ for all $t \in \sR$.
\\

\textbf{Step 4. The general case.} The arguments used in the previous two steps can be easily adapted to any kind of a (nonempty) interval $I$. More concretely, for any $a,b \in \sR$ and $a < b$:
\begin{itemize}
\item For $I = [a,+\infty)$ apply the reasoning from Step 2 with
\begin{align*}
\X_i := C^{[a+i-1,a+i]}_\T\left(\T^{-1}(a+i-1), \T^{-1}(a+i)\right)
\end{align*}
and other minor modifications that are necessary.
\item For $I = [a,b)$ apply the reasoning from Step 2 with
\begin{align*}
\X_i := C^{[b + (a-b) 2^{-i} , b + (a-b) 2^{-i-1}]}_\T\left(\T^{-1}(b + (a-b) 2^{-i}), \T^{-1}(b + (a-b) 2^{-i-1})\right)
\end{align*}
and other minor modifications that are necessary.
\item For $I = (-\infty,b]$ or $I = (-a,b]$ modify the above cases in a similar spirit as Step 3 modified Step 2 in order to construct $\sigma_-$.
\item Finally, for $I = (a,b)$, $I = (a,+\infty)$ or $I = (-\infty,b)$, simply take suitable pairs of $\sigma$'s provided by the previous cases and concatenate them, similarly as in Step 3.
\end{itemize}
\end{Proof}

With Theorem \ref{main} proven, its variant for the spaces $\I_\T$ can be shown to hold without much difficulty.

\begin{Proof} \textbf{of Theorem \ref{mainvar}, $i) \Rightarrow ii)$.}
Theorem \ref{main} guarantees the existence of $\sigma \in \Pf(C^\sR_\T)$ satisfying $(\ev_t)_\# \sigma = \mu_t$ for all $t \in \sR$. Let $\iota: \I_\T \hookrightarrow C^\sR_\T$ denote the canonical topological embedding. Suppose we have shown that $\sigma(\I_\T) = 1$. Then, defining $\upsilon$ via $\upsilon(E) := \sigma(\iota(E))$ for any Borel subset $E \subseteq \I_\T$ one would obtain an element of $\Pf(\I_\T)$ with the desired properties. Indeed, for any $t \in \sR$ and any Borel subset $\X \subseteq \M$ we would have
\begin{align*}
(\ev_t|_{\I_\T})_\# \upsilon(\X) = (\ev_t \circ \iota)_\# \upsilon(\X) = (\ev_t)_\# \upsilon(\iota^{-1}(\X)) = (\ev_t)_\# \sigma(\X) = \mu_t(\X).
\end{align*}

Hence, we only need to show that $\sigma(\I_\T) = 1$. To this end, observe first that
\begin{align}
\label{mainvar1}
\forall \, t \in \sR \quad \sigma\left( (\T \circ \ev_t)^{-1}(t) \right) = (\T \circ \ev_t)_\# \sigma(\{t\}) = \T_\# \mu_t(\{t\}) = \mu_t(\T^{-1}(t)) = 1,
\end{align}
where the last equality follows from $\supp \, \mu_t \subseteq \T^{-1}(t)$.

Remark \ref{ITRem}, the inclusion--exclusion principle and (\ref{mainvar1}) allow to obtain
\begin{align*}
& \sigma(\I_\T) = \sigma\left( (\T \circ \ev_0)^{-1}(0) \cap (\T \circ \ev_1)^{-1}(1) \right)
\\
& = \underbrace{\sigma\left( (\T \circ \ev_0)^{-1}(0) \right)}_{= \, 1} + \underbrace{\sigma\left( (\T \circ \ev_1)^{-1}(1) \right)}_{= \, 1} - \underbrace{\sigma\left( (\T \circ \ev_0)^{-1}(0) \cup (\T \circ \ev_1)^{-1}(1) \right)}_{= \, 1} = 1.
\end{align*}

\textbf{$ii) \Rightarrow i)$} Define $\sigma := \iota_\# \upsilon \in \Pf(C^\sR_\T)$ and observe that for any $t \in \sR$ one has
\begin{align*}
(\ev_t)_\# \sigma = (\ev_t \circ \iota)_\# \upsilon = (\ev_t|_{\I_\T})_\# \upsilon = \mu_t.
\end{align*}
Moreover, $\supp \, \mu_t \subseteq \T^{-1}(t)$ for all $t \in \sR$, because
\begin{align*}
\mu_t(\T^{-1}(t)) = \T_\# \mu_t(\{t\}) = (\T \circ \ev_t|_{\I_\T})_\# \upsilon(\{t\}) = \upsilon\left( (\T \circ \ev_t|_{\I_\T})^{-1}(t) \right) = \upsilon(\I_\T) = 1,
\end{align*}
where the penultimate equality follows from the very definition of $\I_\T$. We can thus apply Theorem \ref{main} and conclude that the map $t \mapsto \mu_t$ must be causal.
\end{Proof}


\section{Appendix: Causality theory}
\label{sec::preliminaries}

For the reader's convenience, in this subsection we recall some basic definitions and facts from causality theory used in above investigations. For a~detailed exposition of this theory the~reader is referred to \cite{Beem,MS08,BN83,Penrose1972} or to Section 2.3 in \cite{EcksteinMiller2015}.

Let $\M$ be a spacetime with metric $g$ (assumed $C^2$). For any $p,q \in \M$, we say that $p$ \emph{causally precedes} $q$, denoted by $p \preceq q$, if there exists a piecewise smooth future-directed causal curve $\gamma: [0,1] \rightarrow \M$ \emph{from $p$ to $q$}, i.e. $\gamma(0) = p$ and $\gamma(1) = q$.

The relation $\preceq$ is reflexive and transitive. If it is additionally antisymmetric, we call $\M$ a \emph{causal} spacetime. In any case, $\prec$ denotes the irreflexive kernel of $\preceq$.

For any $p \in \M$ the sets $J^+(p), J^-(p) \subseteq \M$ denote the \emph{causal future} and \emph{past} of $p$, respectively. One also defines $J^\pm(\X) := \bigcup_{p \in \X} J^\pm(p)$ for any subset $\X \subseteq \M$, whereas $J^+$ \emph{tout court} stands for the set of all pairs $(p,q) \in \M^2$ such that $p \preceq q$.

For the definition of a future-directed causal curve to make sense, one needs the curve to be (piecewise) differentiable. Nevertheless, one can extend this definition to encompass curves which are only continuous \cite{MinguzziCurves08,MS08}. Under a rather mild assumption that $\M$ is a \emph{distinguishing} spacetime (for the definition consult \cite[Section 3.2]{MS08}), this extended definition boils down to the following condition \cite[Prop. 3.19]{MS08}.
\begin{Prop}
\label{futdircaus}
A curve $\gamma \in C(I,\M)$ is future-directed causal iff $\ \forall \, s,t \in I \ \ s < t \ \Rightarrow \ \gamma(s) \prec \gamma(t)$.
\end{Prop}

From now on, every causal curve is assumed future-directed. Following \cite{SanchezProgress}, by a \emph{causal path} we understand the image of a causal curve. In causal spacetimes, a causal path can be equivalently regarded as an equivalence class of causal curves modulo a (continuous and strictly increasing) reparametrization. For this reason, in the paper we adopt the notation $[\gamma]$ to denote the causal path associated to the causal curve $\gamma$. It will always be clear from the context whether one should interpret $[\gamma]$ as a class of curves or as a subset of $\M$.

A~causal curve (path) is called \emph{inextendible} if it has neither a past nor future endpoint. Recall that a~\emph{Cauchy hypersurface} is a~subset $\Sw \subseteq \M$ met exactly once by every inextendible timelike curve. Any such $\Sw$ is a~closed topological hypersurface, met by every inextendible causal curve. Of course, if $\Sw$ is additionally spacelike, then the previous sentence can be strengthened by adding ``exactly once'' at the end.

A~function $\T: \M \rightarrow \mathbb{R}$ is referred to as
\begin{itemize}
\item a~\emph{time function} if it is continuous and strictly increasing along every future-directed causal curve,
\item a~\emph{temporal function} if it is a~smooth function with a past-directed timelike gradient.
\end{itemize}
Although ``temporal'' implies ``time'' function, even a smooth time function need not be temporal.

We now recall three more definitions concerning the causal properties of a spacetime $\M$, each of them stronger than the~preceding one.

$\M$ is called \emph{stably causal} if it admits a time function. Every stably causal spacetime is distinguishing and causal. Moreover, it admits a temporal function as well.

$\M$ is called \emph{causally simple} if it is causal and satisfies one of the~following equivalent conditions \cite[Proposition 3.68]{MS08}:
\begin{itemize}
\item $J^+(p)$ and $J^-(p)$ are closed for every $p \in \M$.
\item $J^+(\K)$ and $J^-(\K)$ are closed for every compact $\K \subseteq \M$.
\item $J^+$ is a~closed subset of $\M^2$.
\end{itemize}

Finally, $\M$ is called \emph{globally hyperbolic} if it satisfies one of the~following equivalent conditions:
\begin{itemize}
\item $\M$ is causal and the~sets $J^+(p) \cap J^-(q)$ are compact for all $p,q \in \M$.
\item $\M$ admits a temporal function $\T$, the~level sets of which are (smooth spacelike) Cauchy hypersurfaces \cite{BS04}.
\end{itemize}
Any time (temporal) function whose level sets are Cauchy hypersurfaces is called a \emph{Cauchy} time (temporal) function.

\begin{Prop}
\label{PropCCC}
Let $\M$ be a globally hyperbolic spacetime. The following subsets of $\M$ are compact:
\begin{enumerate}[\itshape i)]
\item $J^+(\K_1) \cap J^-(\K_2)$ for any compact $\K_1, \K_2 \subseteq \M$,
\item $J^\pm(\K) \cap \Sw$ for any compact $\K \subseteq \M$ and any Cauchy hypersurface $\Sw \subseteq \M$,
\item $J^\pm(\K) \cap J^\mp(\Sw)$ for any compact $\K \subseteq \M$ and any Cauchy hypersurface $\Sw \subseteq \M$.
\end{enumerate}
\end{Prop}
\begin{Proof}\textbf{.}
For the proofs of \emph{i), ii)} see \cite[Lemma 11.5]{HR09} and \cite[Property 4 on p. 44]{MS08}, respectively. Here, let us only show the compactness of $J^+(\K) \cap J^-(\Sw)$ (the proof for $J^-(\K) \cap J^+(\Sw)$ is analogous).

Observe that $J^+(\K) \cap J^-(\Sw) = J^+(\K) \cap J^-(\Sw \cap J^+(\K))$. Indeed, the inclusion ``$\supseteq$'' is obvious, whereas in order to prove ``$\subseteq$'' suppose $r \in J^+(\K) \cap J^-(\Sw)$. It means there exist $p \in \K$ and $q \in \Sw$ such that $p \preceq r \preceq q$. But then $p \preceq q$ and so $q$ in fact belongs to $\Sw \cap J^+(\K)$.

Notice now that the latter set is compact by \emph{ii)} and hence, on the strength of \emph{i)}, the intersection $J^+(\K) \cap J^-(\Sw \cap J^+(\K))$ is compact as well.
\end{Proof}

In their seminal paper \cite{BS04}, Bernal and S\'{a}nchez proved the smooth version of Geroch's splitting theorem \cite{Geroch1970}. Their method was to construct a Cauchy temporal function in any given globally hyperbolic spacetime, because with any such a function $\T$ one can always create a smooth splitting ``associated'' to $\T$ in the sense of the following theorem (see \cite{BS04} or \cite[Remark 3.4]{BSProc}).
\begin{Thm}[(Geroch, Bernal, S\'{a}nchez)]
\label{GBS}
Let $(\M,g)$ be a globally hyperbolic spacetime with metric $g$ and let $\T: \M \rightarrow \sR$ be a Cauchy temporal function. Then there exists an isometry $\Phi: \M \rightarrow \sR \times \itSigma$, called here the \emph{Geroch--Bernal--S\'{a}nchez} (GBS) \emph{splitting}, where $\itSigma := \T^{-1}(0)$, $\T = \Phi^\ast \pi^1$ such that
\begin{align}
\label{Geroch}
(\Phi^\ast)^{-1} g = - \beta \, d\pi^1 \otimes d\pi^1 + \mathcal{G}
\end{align}
\noindent
with $\beta: \sR \times \itSigma \rightarrow \sR$ a positive smooth function and $\mathcal{G}$ a 2-covariant symmetric tensor field on $\sR \times \itSigma$, whose restriction to $\{t\} \times \itSigma$ is a Riemannian metric for every $t \in \sR$ and whose radical at each $(t,x) \in \sR \times \itSigma$ is spanned by the gradient $(\grad \, \pi^1)|_{(t,x)}$.
\end{Thm}

Let us pull formula (\ref{Geroch}) back on $\M$. Applying $\Phi^\ast$ to both sides, we can write that
\begin{align}
\label{Geroch2}
g = - \alpha \, d\T \otimes d\T + \bar{g},
\end{align}
\noindent
where $\alpha := \Phi^\ast \beta$ is a positive smooth function on $\M$ and $\bar{g} := \Phi^\ast \mathcal{G}$ is a 2-covariant symmetric tensor field on $\M$, whose restriction to $\Phi^{-1}\left(\{t\} \times \itSigma\right) = \T^{-1}(t)$ is a Riemannian metric for every $t \in \sR$ and whose radical at each $p \in \M$ is spanned by the gradient $(\grad \, \T)|_p$.

Consider the ``corresponding'' Riemannian metric on $\M$:
\begin{align}
\label{Rmetric}
w_0 := \alpha \, d\T \otimes d\T + \bar{g}.
\end{align}
By the celebrated result of Nomizu and Ozeki \cite{NomizuOzeki}, there exists a smooth positive function $u$ on $\M$ such that $w := u w_0$ is a \emph{complete} Riemannian metric. Needless to say, its associated distance function $d_w$ induces the original manifold topology of $\M$. Observe that $w$ and $g$ are related through
\begin{align}
\label{Rmetric2}
w = ug + 2u \alpha \, d\T \otimes d\T.
\end{align}

\section*{Acknowledgments}
The author wishes to thank Micha{\l} Eckstein for the careful reading of the manuscript as well as for all enlightening discussions and helpful comments.

This work was supported by a grant from the John Templeton Foundation (grant ID\# 60671).

\bibliographystyle{plain}
\bibliography{causality}{}

\end{document}